\documentclass{JINST}
\title{Prospects for measuring the gravitational free-fall of antihydrogen with  emulsion detectors\\ 
\bigskip\hspace{5cm}\small AEgIS Collaboration}

\author{S.~  Aghion,$^{a,b}$  
O.~Ahl\'en,$^c$  
C.~  Amsler,$^d$\thanks{Corresponding author.}\,
A.~  Ariga,$^d$
T.~  Ariga,$^d$ 
A.~  S.~  Belov,$^e$  
G.~  Bonomi,$^{f,g}$ 
P.~  Br\"aunig,$^h$ 
J.~  Bremer,$^c$ 
R.~  S.~  Brusa,$^i$ 
L.~ Cabaret,$^j$  
C.~  Canali,$^k$ 
R.~  Caravita,$^l$
F.~  Castelli,$^l$ 
G.~  Cerchiari,$^l$
S.~  Cialdi,$^l$ 
D.~  Comparat,$^j$ 
G.~  Consolati,$^{m}$
J.~ H.~ Derking,$^c$
S.~  Di~ Domizio,$^{n,o}$ 
L.~  Di~ Noto,$^i$ 
M.~  Doser,$^c$ 
A.~  Dudarev,$^c$ 
A.~  Ereditato,$^d$  
R.~  Ferragut,$^{a,b}$ 
A.~  Fontana,$^g$ 
P.~  Genova,$^g$
M.~  Giammarchi,$^b$ 
A.~  Gligorova,$^p$ 
S.~ N.~ Gninenko,$^e$ 
S.~  Haider,$^c$ 
J.~ Harasimovicz,$^q$
S.~  D.~  Hogan,$^r$ 
T.~  Huse,$^s$ 
E.~  Jordan,$^t$ 
L.~  V.~  J\o rgensen,$^c$ 
T.~  Kaltenbacher,$^c$  
J.~  Kawada,$^d$ 
A.~  Kellerbauer,$^t$  
M.~  Kimura,$^d$  
A.~Knecht,$^c$ 
D.~ Krasnick\'y,$^{n,o}$  
V.~  Lagomarsino,$^{n,o}$
A.~Magnani,$^g$
S.~  Mariazzi,$^i$ 
V.~  A.~  Matveev,$^{e,u}$ 
F.~  Moia,$^{a,b}$ 
G.~  Nebbia,$^v$ 
P.~  N\'ed\'elec,$^w$ 
M.~ K.~ Oberthaler,$^h$ 
N.~  Pacifico,$^p$ 
V.~ Petr\'a\v{c}ek,$^x$  
C.~  Pistillo,$^d$  
F.~  Prelz,$^b$ 
M.~  Prevedelli,$^y$ 
C.~  Regenfus,$^k$ 
C.~  Riccardi,$^{g,z}$ 
O.~  R\o hne,$^s$ 
A.~  Rotondi,$^{g,z}$ 
H.~  Sandaker,$^p$ 
P.~  Scampoli,$^{d,zz}$ 
A.~ Sosa,$^q$
J.~  Storey,$^d$ 
M.~ A.~  Subieta Vasquez,$^{f,g}$ 
M.~  \v{S}pa\v{c}ek,$^x$ 
G.~  Testera,$^n$ 
D.~  Trezzi,$^b$ 
R.~  Vaccarone,$^n$ 
C.~ P.~ Welsch,$^q$
S.~  Zavatarelli$^n$ \\
\llap{$^a$} Politecnico di Milano, Piazza Leonardo da Vinci 32, 20133 Milano, Italy\\
\llap{$^b$} Istituto Nazionale di Fisica Nucleare, Sez. di Milano, Via Celoria 16, 20133 Milano, Italy\\
\llap{$^c$} European Organisation for Nuclear Research, Physics Dept., 1211 Geneva 23, Switzerland\\
\llap{$^d$} Albert Einstein Center for Fundamental Physics, Laboratory for High Energy Physics, 
University of Bern, 3012 Bern, Switzerland\\
\llap{$^e$} Institute for Nuclear Research of the Russian Academy of Sciences, Moscow 117312, Russia\\
\llap{$^f$} University of Brescia, Dept. of Mechanical and Industrial Engineering, Via Branze 38, 25133 Brescia, Italy\\
\llap{$^g$} Istituto Nazionale di Fisica Nucleare, Sez. di Pavia, Via Agostino Bassi 6, 27100 Pavia, Italy\\
\llap{$^h$} University of Heidelberg, Kirchhoff Institute for Physics, Im Neuenheimer Feld 227, 69120 Heidelberg, Germany\\
\llap{$^i$} Dipartimento di Fisica, Universit\`a di Trento and INFN, Gruppo Collegato di Trento, Via Sommarive 14, 38123 Povo, Trento, Italy\\
\llap{$^j$} Laboratoire Aim\'e Cotton, CNRS, Universit\'e Paris Sud, ENS Cachan, B\^atiment 505, Campus d'Orsay, 91405 Orsay Cedex, France\\
\llap{$^k$} University of Zurich, Physics Institute, Winterthurerstrasse 190, 8057 Zurich, Switzerland\\
\llap{$^l$} University of Milano, Dept. of Physics, Via Celoria 16, 20133 Milano, Italy\\
\llap{$^{m}$} Dept. of Aerosp. Sci. and Tech., Politecnico di Milano, via La Masa, 34 20156 Milano, Italy\\
\llap{$^n$} Istituto Nazionale di Fisica Nucleare, Sez. di Genova, Via Dodecaneso 33, 16146 Genova, Italy\\
\llap{$^o$} University of Genoa, Dept. of Physics, Via Dodecaneso 33, 16146 Genova, Italy\\
\llap{$^p$} University of Bergen, Institute of Physics and Technology, Alleegaten 55, 5007 Bergen, Norway\\
\llap{$^q$} University of Liverpool and Cockroft Institute, UK\\
\llap{$^r$} University College London, Dept. of Physics and Astronomy, Gower Street, London WC1E 6BT,UK\\
\llap{$^s$} University of Oslo, Dept. of Physics, Sem S\ae lands vei 24, 0371 Oslo, Norway\\
\llap{$^t$} Max Planck Institute for Nuclear Physics, Saupfercheckweg 1, 69117 Heidelberg, Germany\\
\llap{$^u$} Joint Institute for Nuclear Research, 141980 Dubna, Russia\\
\llap{$^v$} Istituto Nazionale di Fisica Nucleare, Sez. di Padova, Via Marzolo 8, 35131 Padova, Italy\\
\llap{$^w$} Claude Bernard University Lyon 1, Institut de Physique Nucl\'eaire de Lyon, 4 Rue Enrico Fermi, 69622 Villeurbanne, France\\
\llap{$^x$} Czech Technical University in Prague, FNSPE, B\v{r}ehov\'a 7, 11519 Praha 1, Czech Republic\\
\llap{$^y$} University of Bologna, Dept. of Physics, Via Irnerio 46, 40126 Bologna, Italy\\
\llap{$^z$} University of Pavia, Dept. of Nuclear and Theoretical Physics, Via Bassi 6, 27100 Pavia, Italy\\
\llap{$^{zz}$} University of Napoli Federico II, Dept. of Physics, Via Cinthia, 80126 Napoli, Italy\\
E-mail: \email{claude.amsler@cern.ch}}

\abstract{The main goal of the AEgIS experiment at CERN is to test the weak equivalence principle for antimatter. AEgIS will measure the free-fall of an antihydrogen beam traversing a moir\'e deflectometer. The goal is to determine the gravitational acceleration $\bar{g}$ with an initial relative accuracy of 1\% by using an emulsion detector combined with a silicon $\mu$-strip  detector to measure the time of flight. Nuclear emulsions can measure the annihilation vertex of antihydrogen atoms with a precision of $\sim$1 -- 2 $\mu$m r.m.s. We present here results for emulsion detectors operated in vacuum using low energy antiprotons from the CERN antiproton decelerator. We  compare with Monte Carlo simulations, and discuss the impact on the  AEgIS project.}

\begin{document}

\section{Introduction}
We intend to use  nuclear emulsions to measure the  gravitational sag of an antihydrogen beam at the CERN antiproton decelerator (AD). In the AEgIS experiment (AD6) the free-fall of antihydrogen atoms launched horizontally will be measured to directly determine for the first time the gravitational acceleration of antimatter by matter. The vertical precision on the measured annihilation point will be around 1 -- 2 $\mu$m r.m.s. This will be achieved by adding an emulsion detector to the originally foreseen $\mu$-strip detector \cite{propaegis}. For the first time nuclear emulsion films will be used in vacuum. 

According to the weak equivalence principle (WEP) the trajectory of a point mass in the presence of gravity does not depend on its composition, but only on the two kinematical parameters of initial position and velocity. This principle of the universality of free fall, a cornerstone of General Relativity, has been tested with  precisions down to  10$^{-13}$ by Etv\"os-type experiments using probes made of various materials \cite{Adelberger}. However, the WEP has never been tested with antimatter particles, since electromagnetic disturbances could not be reduced to a level such that the effect of the much weaker gravitational force could be observed. In contrast, the antihydrogen atom is electrically neutral and as such an ideal probe to test the WEP. 

Many theoretical arguments have been proposed to rule out any difference between the acceleration of matter $g$ (9.81 ms$^{-2}$ at sea level) and that of antimatter $\bar{g}$ and, correspondingly,  many rebuttals have been published \cite{Goldman} (for a more recent review see  ref. \cite{Fischler}). Predictions for the relative difference $\Delta g/g$ between 200\% (antigravity with $\bar{g} = - g$) and less than 10$^{-13}$ can be found in the literature, based on indirect measurements, and depending on theoretical assumptions. 

In quantum field theories the exchange of graviscalar bosons S and tensor gravitons T is attractive for matter-matter, while the exchange of graviphotons V is repulsive for matter-matter and attractive for matter-antimatter gravitation. Thus attractive coherence between S and V  is expected in antimatter-matter while partial cancellation is expected in matter-matter, leading to $\bar{g} > g$. However, recent torsion balance experiments searching for a 5$^{\rm th}$ force with a sensitivity better than 10$^{-13}$ \cite{Adelberger}, and which are sensitive to V but not S, rule out any significant V contribution in matter-matter, unless there is a nearly perfect cancellation of V by S \cite{Adelantimatter}. Precise cancellations are deemed to be possible \cite{Goldman} or, if excluded, torsion balance experiments lead to $\Delta g/g< 10^{-7}$ \cite{Alves}. In any case these arguments all rest on CPT which has not been tested for gravity.

Another argument against $g\neq\bar{g}$ is the tiny upper limit for the mass difference in the $K^0-\bar{K}^0$ system. Assuming CPT invariance (which is established at the level of 8 $\times$ 10$^{-19}$ {\it in this system}), ref. \cite{Karshen1} arrives at an upper limit of $\Delta g/g = 8 \times 10^{-13}$. However, the neutral kaons  are a mixture of matter-antimatter ($d\bar{s}$, resp. $\bar{d}s$) and {\it are not baryons}. Hence cancellations could happen for mesons, but not for baryons. In fact,  the standard model extension  of ref. \cite{Kostel} - based on field theories - contain tiny CPT flavour dependent violations which could become apparent in baryons, while remaining hidden in mesons.

Attempts are being made to  measure the 1s -- 2s transition frequency in positronium with high precision \cite{Cassidy}. The frequency, which is redshifted on the earth due to the gravity of the sun,  can be predicted by using the well measured value for hydrogen. The measured frequency \cite{Fee} is somewhat {\it below} the predicted value, while gravitational repulsion on the positron would blueshift the frequency {\it above} the predicted value.  This makes antigravity ($\bar{g}=-g$) unlikely. 

The issue of CPT and the validity of WEP for antimatter eventually rests on direct measurements.  An experimental method to address the question of the gravitational nature of antimatter has been tested  recently at the CERN-AD \cite{Alpha}.
The AEgIS experiment \cite{propaegis}, also at the CERN-AD (AD6), will be the first direct measurement of $\bar{g}$, starting  in 2014. Our goal is to achieve an initial precision of 1\%  on $\Delta g/g$ and to pave the way for future improvements. Indeed, any precision would be valuable, while we propose here to perform the best possible measurement with advanced technologies. Another experiment, GBAR \cite{GBAR}, will be installed on the ELENA very low energy antiproton ring at the CERN-AD, due to start operation in 2018. The latter will attempt to generate very cold antihydrogen by trapping $\overline{\textrm{H}}^+$-atoms, with the final goal of reaching a  precision of 0.1\% on $\bar{g}$. 

\section{The AEgIS experiment}
\subsection{Overview of the apparatus}
\begin{figure}[htb]
\parbox{75mm}{\mbox{
\includegraphics[width=70mm]{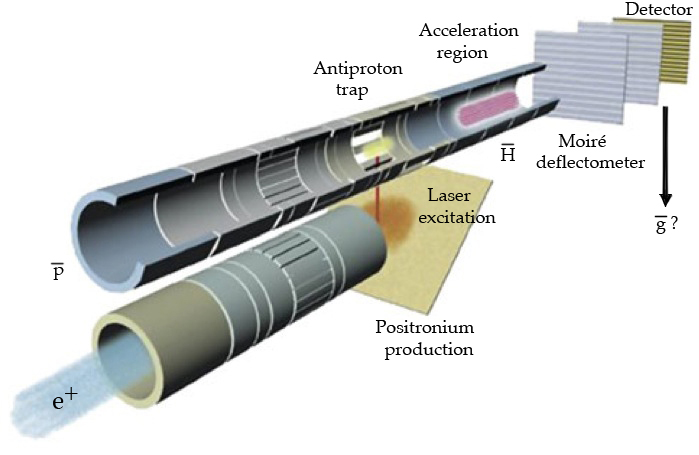}
}\centering}\hfill
\parbox{75mm}{\mbox{
\includegraphics[width=70mm]{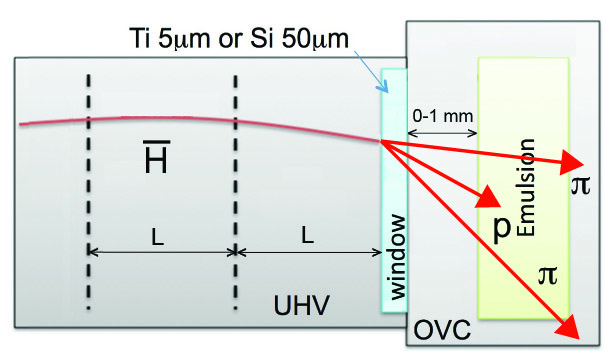}
}\centering} \caption[]{\it Left: sketch of the AEgIS apparatus. The $\overline{\textrm{H}}$-atoms are produced in the $\bar{p}$-trap in a 1 T field. Not shown are the $\bar{p}$ and e$^+$ capture traps in the 5 T magnet at the entrance of the apparatus. Right: Schematics of the AEgIS detectors. The vertex detector is made of a thin annihilation foil  followed by nuclear emulsions. A time of flight  detector (not shown) is needed to measure the velocity of the $\overline{\textrm{H}}$ atoms. \label{AEGISPrinz}}
\end{figure}

AEgIS  \cite{propaegis} first needs to produce very low energy $\overline{\textrm{H}}$-atoms. Figure~\ref{AEGISPrinz} (left) shows a sketch of the apparatus. The process begins with the production of positronium (Ps) by accelerating $10^{8}$ positrons from a positron accumulator \cite{Greaves} onto a nanoporous material \cite{Mariazzi}. The ortho-Ps emitted from the target is then brought to the Rydberg state Ps* by  two-step laser excitation. Some of the Ps* atoms drift into a Penning trap in which $10^{5}$ antiprotons ($\bar{\textrm{p}}$)  have been stored and cooled to sub-K temperature, producing $\overline{\textrm{H}}$ through the charge exchange reaction $\textrm{Ps}^{*} + \overline{\textrm{p}} \rightarrow \overline{\textrm{H}}^{*} + \textrm{e}^{-}$.  This reaction between the antiproton and the highly excited Rydberg state positronium (Ps*), in which the bound positron is captured by the antiproton and an electron is released, was  proposed in ref. \cite{Charlton}. The $\overline{\textrm{H}}$-atoms follow a Maxwell-Boltzmann distribution with an average speed of $\sim$50 m/s. An electric field is then applied to accelerate the Rydberg $\overline{\textrm{H}}$-atoms  to $\sim$400 m/s \cite{GemmaOki}. This technique has   been demonstrated with hydrogen atoms \cite{Merkt}. 

The measurement principle is sketched in figure~\ref{AEGISPrinz} (right).  A moir\'{e} deflectometer \cite{Oberthaler} consisting of two identical gratings (separated by a distance $L$ of $\sim$50 cm) and a thin foil located at the same distance $L$ from the second grating are used to measure the deflection of the $\overline{\textrm{H}}$-beam in the  gravitational field. The vertical distribution of  annihilation vertices on the foil (moir\'e intensity pattern) will be recorded with an r.m.s. resolution of $\sigma$ $\sim$1 -- 2 $\mu$m. This will be achieved by employing nuclear emulsion films to track annihilation products, as detailed in the following.  The thin foil, typically 50 $\mu$m (active) silicon or 5 $\mu$m titanium, is needed as a window to separate the vacuum of the inner cryogenic UHV chamber from the insulating outer vacuum chamber (OVC) containing the emulsion detector.

Due to gravity the moir\'e  intensity pattern will be shifted when compared to the one obtained e.g. with a light source. The vertical displacement $\Delta x$ of the moir\'e  pattern  can be measured from the time of flight $T$  between the two gratings through the relation $\Delta x$ = $\bar{g} T^2$. 
Thus a segmented time of flight detector (not shown in  figure \ref{AEGISPrinz}) is needed to measure the velocity of the atoms, and to record the approximate position of the annihilation vertex for subsequent offline analysis with only a moderate precision of about 1 mm. This can be achieved either by using as window a thin  $\mu$-strip detector separated by a gap of a few 100 $\mu$m from the emulsion detector, or with a silicon or scintillating fiber tracker located behind the emulsion detector, in which case the window would consist of a $\sim$5 $\mu$m titanium foil. The time of flight can be computed from the switch off time of the electric field for Stark acceleration and the $\overline{\textrm{H}}$ annihilation time. 

The antiproton beam will be injected into the apparatus every $\sim$100 s with a shot of typically 3 $\times$ 10$^7$ $\bar{\textrm{p}}$. The antiprotons that are not trapped are stopped in the center of the apparatus,  350 cm upstream of the emulsions. This  would give an estimated background rate of $\sim$300 tracks/cm$^2$ in the emulsions for every $\bar{\textrm{p}}$ shot from the AD. During the 100 s between  shots AEgIS expects to produce about  10 $\overline{\textrm{H}}$-atoms and to accelerate them to average velocities of 400 m/s. The number of $\overline{\textrm{H}}$-atoms reaching the annihilation detector depends on the exact geometry of the free-fall section, which has not been finalized yet.  With the current design (simulated in section \ref{sec:42}) we estimate that roughly 3\% of the $\overline{\textrm{H}}$-atoms will reach the end of the launching tube, since the majority will annihilate on the walls or gratings. Incidentally, this annihilation background will  hardly affect the measurement since the emulsion detector will determine the annihilation point.  

We will use the scanning facility  available at the Laboratory for High Energy Physics (LHEP) in Bern, which was developed for the OPERA experiment \cite{OPERA}. The emulsions can be left in the final apparatus for several days before being replaced, since the maximum track density that can easily be dealt with our current scanning facility at LHEP  is around 10$^5$ tracks/cm$^2$. 

The absolute alignment between deflectometer and emulsion detector can be controlled  with  collimated $^{55}$Fe-sources \cite{Aoki} or with a light source.  Local expansion can be corrected by printing a reference frame with sub-micrometric accuracy (photomask) on the emulsion films \cite{Emul,Kimura}. Since $\Delta x$ depends on the $\overline{\textrm{H}}$ velocity, fast $\overline{\textrm{H}}$-atoms can also be used to determine the vertical alignment of the apparatus components. 

\subsection{Emulsion detectors} 
Emulsion detectors can be used as tracker devices to determine the $\overline{\textrm{H}}$ annihilation vertices with micrometer    resolutions, or even  as targets by directly observing the annihilation vertices. For a general review on emulsion detectors see ref. \cite{Lellis}. In AEgIS the deflectometer and detector will be operated  in vacuum and at temperatures between 77 K and room temperature.  Accordingly, an intensive R\&D programme on emulsion films has been started to cope with their operation under such conditions. First results on the behaviour of emulsions in vacuum have already been published \cite{Emul}. Here we report on results achieved with antiprotons annihilating in emulsions in vacuum and at room temperature (see also ref. \cite{Vienna}). 

We are also investigating new emulsion gels with higher sensitivity to increase the detection efficiency for minimum ionizing particles (MIPs) and using glass instead of plastic as the substrate. Glass is well suited for highest position resolutions thanks to its superior environmental stability (temperature and humidity), as compared to plastic. In the following we shall refer to the original emulsion gel produced by FUJIFILM on plastic (triacetyl cellulose, TAC) and used for the OPERA experiment as the {\it standard} gel \cite{Nakamura}, while the new product with higher sensitivity and glass substrates will be referred to as the {\it new} gel. Emulsions of the new type were manufactured at LHEP with gel provided by the University of Nagoya. 

\begin{figure}[htb]
\parbox{75mm}{\mbox{
\includegraphics[width=70mm]{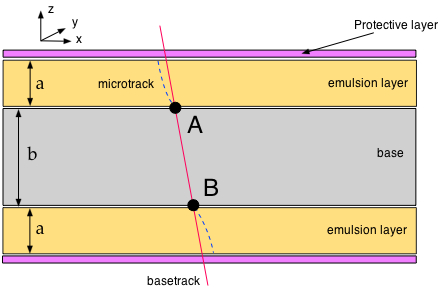}
}\centering}\hfill
\parbox{75mm}{\mbox{
\includegraphics[width=70mm]{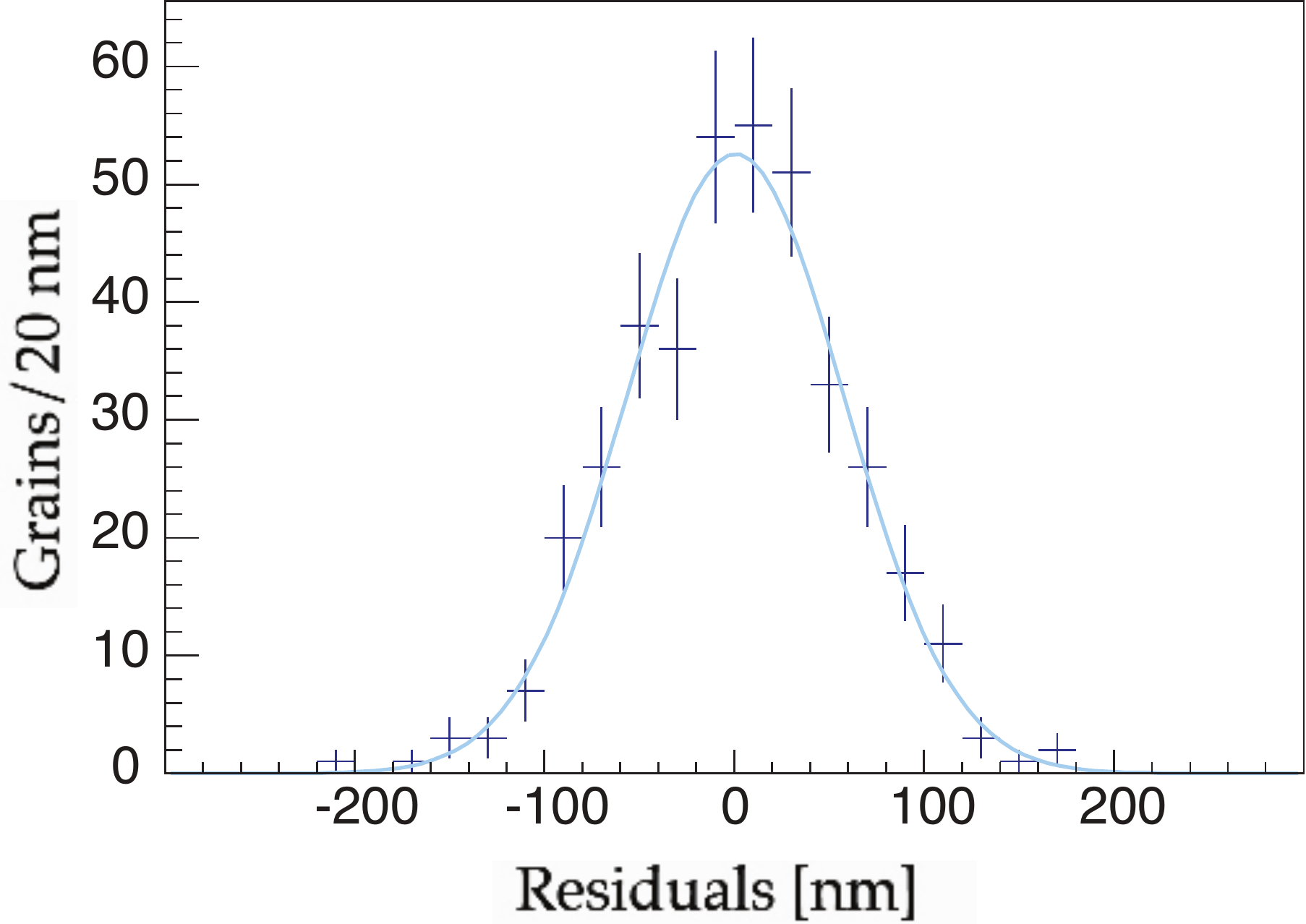}
}\centering} \caption[]{\it Left: sketch of an emulsion film (see text). Right: distribution of the distance between grains and a straight line fit for the new gel (for a track perpendicular to the emulsion surface). \label{Glossary}}
\end{figure}

Let us first introduce a brief glossary (figure~\ref{Glossary}, left).   A typical {\it emulsion detector} is composed of two sensitive emulsion {\it layers} coated on both sides of a substrate or {\it base}. A protective gelatin layer ($\sim$1 $\mu$m thick) covers the outer  surfaces. The detection units are AgBr(I) crystals with  diameters of typically 0.2 $\mu$m which are distributed homogeneously in  gelatin  and which, once excited by a charged particle, form a {\it latent} image. The developing treatment produces filaments of pure Ag  which appear as $\sim$0.6 $\mu$m diameter {\it grains} under the microscope. The {\it fog density} is the density of thermally induced grains. 

Digitalization is performed with an optical microscope. The automatic microscope performs at each point $(x, y)$ a tomographic scan through the two layers ($z$-direction, see figure~\ref{Glossary}). Each layer is scanned in steps of typically 3 $\mu$m in the $z$-direction, leading to 20 tomographic images -- or {\it frames} -- captured by a  camera and sent to a PC. The collection of frames is called a {\it view}. The size of the view depends on the depth of field of the lens and typically  covers a surface of  300 $\times$ 400 $\mu$m$^2$. The tracking algorithm searches for aligned sequences of clusters of grains belonging to different frames of the same view. A linear fit is then performed in the projections $x-z$ and $y-z$ to define the position and angle of the {\it microtrack} which crosses the emulsion layer. Due to distortions at the time of development the microtrack cannot be used directly. Instead, a search is made for a matching microtrack  in the opposite layer and a {\it basetrack} is defined by the the straight line joining points A and B (see figure~\ref{Glossary}). The basetrack is then used as the particle track. 

We have studied in the laboratory  the properties of emulsions in vacuum at a pressure of $\sim$10$^{-6}$ mbar \cite{Emul}. We have treated emulsion films with glycerin solutions of various concentrations to prevent the appearance of cracks due to elasticity loss.  At low glycerin content the fog density reaches unacceptably high values. We therefore settled for the present test to a 12\%  glycerin concentration for the standard  gel for which the fog density reaches a stable value, roughly twice that given in table \ref{table:properties} (18 grains/1000 $\mu$m$^3$). Tests on film sensitivity were conducted on a 6 GeV pion beam at CERN \cite{Emul}. The films were treated with various glycerin contents (4, 8, 12 and 17\%), for 60 minutes, and dried for 1 day at a relative humidity of about 60\%. We found that the addition of glycerin does not affect noticeably the detection efficiency  ($\sim$13\% per AgBr crystal for a MIP in the standard gel). For the new gel we used a  glycerin concentration of 1.5\%.

\begin{table}[htb]
\begin{center}
\begin{tabular}{cccccccc}
\hline
Gel& Density & Layer  & Base  & Crystal & Grain   &Volume & Fog  \\
& & thickness & thickness & size & density & occupancy & density\\
             \hline
                & g$\cdot$cm$^{-3}$ &  $a$ [$\mu$m] &  $b$ [$\mu$m] & [$\mu$m]  & [grains & [\%] & [grains  \\
                &&& &&/100 $\mu$m]&&/1000 $\mu$m$^3$]\\ 
                \hline
Standard   & 2.7 & 44 & 205  & 0.2 & 30.3 $\pm$ 1.6 &30 &  $10.1~\pm$ 0.7 \\
&&&(plastic)&&&&\\
New   & 4.1$^\dagger$& 90 & 1600 & 0.2 &  55.1 $\pm$ 2.6 &55  & $~~3.0~\pm$ 0.3 \\
&&&(glass)&&&&\\ 
\hline
\end{tabular}
\caption[]{\it Characteristics of the gels used during the tests. $^\dagger$Calculated density.}
\label{table:properties}
\end{center}
\end{table}

The intrinsic spatial resolution of emulsions made of the new gel is  found to be $\sigma\sim$0.058 $\mu$m (figure~\ref{Glossary}, right), similar to that of the standard one. Table \ref{table:properties} summarizes the characteristics of the gels used in our tests and table \ref{table:composition}  the composition of the standard gel. The composition of the new gel is similar, but the volume occupancy of the AgBr(I) crystals is larger, as stated in table \ref{table:properties}.

\begin{table}[htb]
\begin{center}
\begin{tabular}{l l l l l l l}
\hline
Ag & Br & I & C & N & O & H   \\
\hline
38.3 & 27.9 & 0.8 & 13.0 & 4.8 &12.4 & 2.4\\
\hline
\end{tabular}
\caption[]{\it Atomic composition  of the standard gel (in \%) apart from traces of other elements such as I, Na, S and Si \cite{Nakathesis}.}
\label{table:composition}
\end{center}
\end{table}

\section{Measurements with antiprotons}
\begin{figure}[htb]
\parbox{75mm}{\mbox{
\includegraphics[width=75mm]{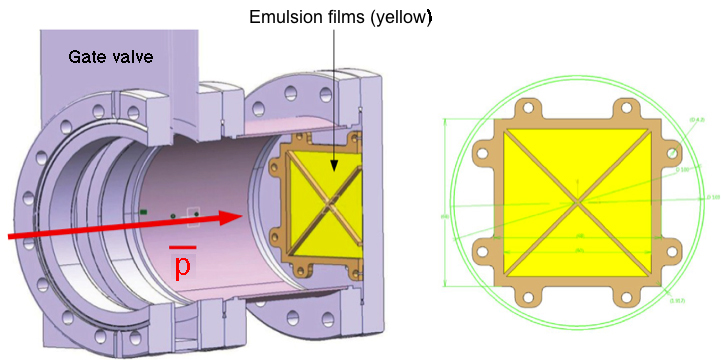}
}\centering}\hfill
\parbox{75mm}{\mbox{
\includegraphics[width=75mm]{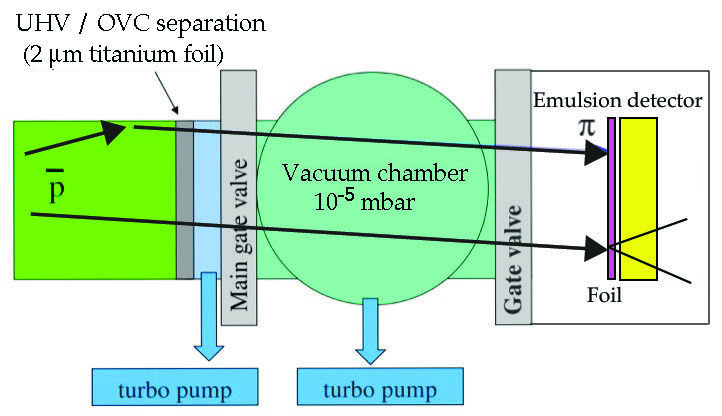}
}\centering} \caption[]{\it Left: vacuum chamber containing the emulsion detector.  Right: beam line setup. Secondary particles (mainly pions) are generated by antiprotons annihilating in the upstream part of the AEgIS apparatus or directly in the emulsion detector. \label{Testsetup}}
\end{figure}

In 2012 we performed tests of emulsions in the AEgIS antiproton beam line at the CERN-AD. The goal was to (i) investigate the performance of emulsions in vacuum, (ii) measure the background rates, and (iii) determine the resolution on the annihilation vertex. The setup is shown in figure~\ref{Testsetup}. The AD produced 5.3 MeV antiprotons (3 $\times$ 10$^7$ $\bar{\textrm{p}}$ / shot every 100 s) which were degraded in an absorber to maximize the capture efficiency in the  antiproton capture trap. We used the antiprotons that passed the HV barrier of the capture trap. The average energy of the antiprotons reaching the emulsion detector at the end of the apparatus was about 100 keV, which corresponds to a range of $\sim$1 $\mu$m in the gel. Therefore the antiprotons annihilated at the surface of the detector. 

\begin{figure}[htb]
\parbox{150mm}{\mbox{
\includegraphics[width=50mm]{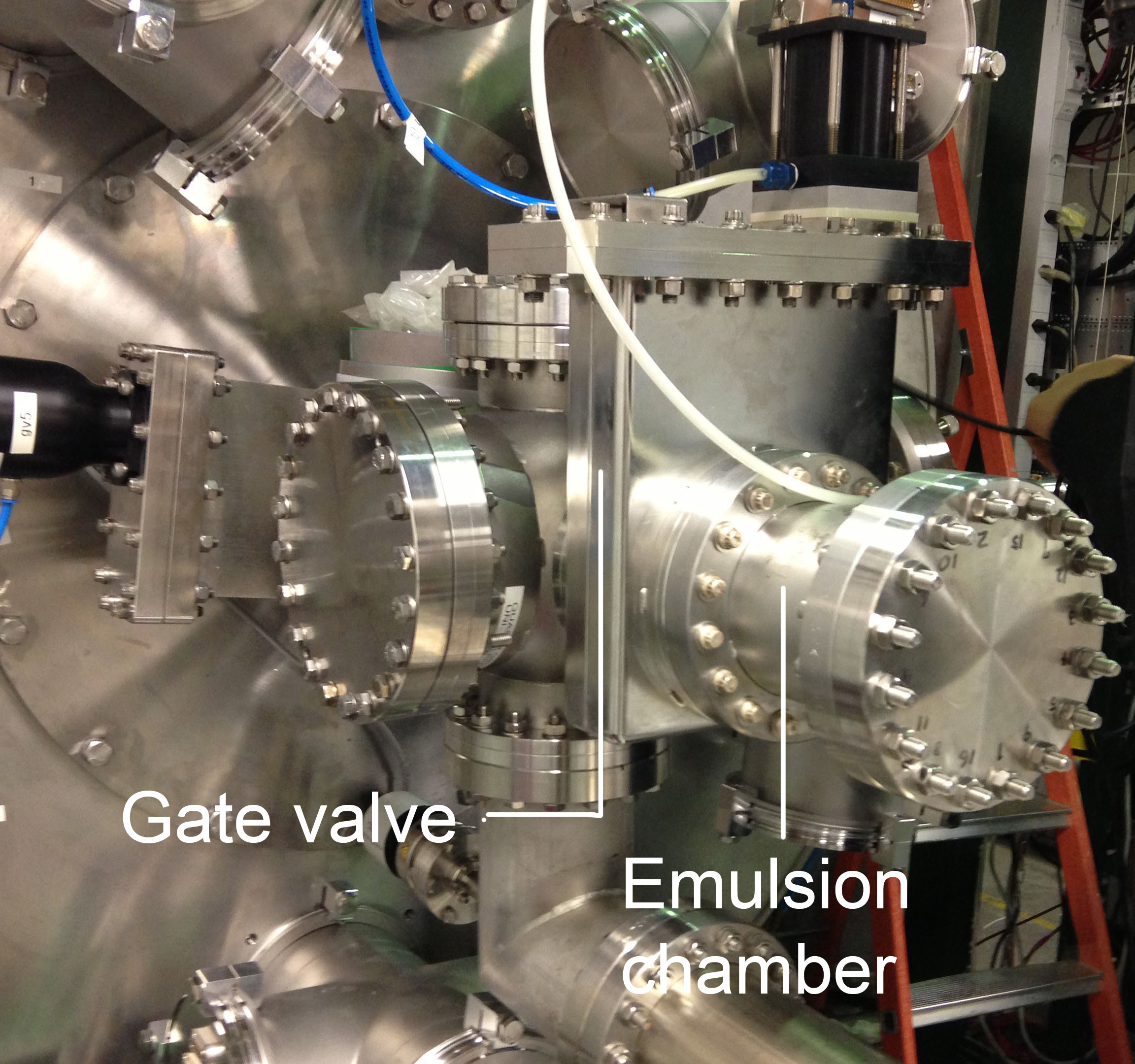}
}\centering}\hfill
\caption[]{\it  Photograph of the emulsion chamber at the downstram end of the AEgIS apparatus.
\label{Photogate}}
\end{figure}

The annihilation detector was installed in a small vacuum chamber shown on the left of figure~\ref{Testsetup}.  The chamber had an inner diameter of 100 mm and the distance from the gate valve to the detector was 114 mm.  The detector itself consisted of a stack of five standard emulsion detectors, with some tests performed with  metallic foils mounted on the surface of the first detector. Some of the runs were also performed with one detector made of emulsion layers of the new type. The emulsion films (68 $\times$ 68 mm$^2$) were held by a stainless steel frame (figure~\ref{Testsetup}, middle), known  to be chemically compatible with emulsions.  The chamber was closed with a gate valve for mounting and dismounting without breaking the AEgIS vacuum, and to protect the films from light exposure. A photograph of the emulsion chamber is shown in figure~\ref{Photogate}.

The emulsion films were then analyzed with the automatic microscope facility at the LHEP. For the standard gel the track reconstruction efficiency was typically 80\% per emulsion. Measurements were also performed with the beam off to measure the muon  flux  from the antiproton production target.  No tracks were found, thus showing that this background is negligible.

\begin{figure}[htb]
\parbox{90mm}{\mbox{
\includegraphics[width=90mm]{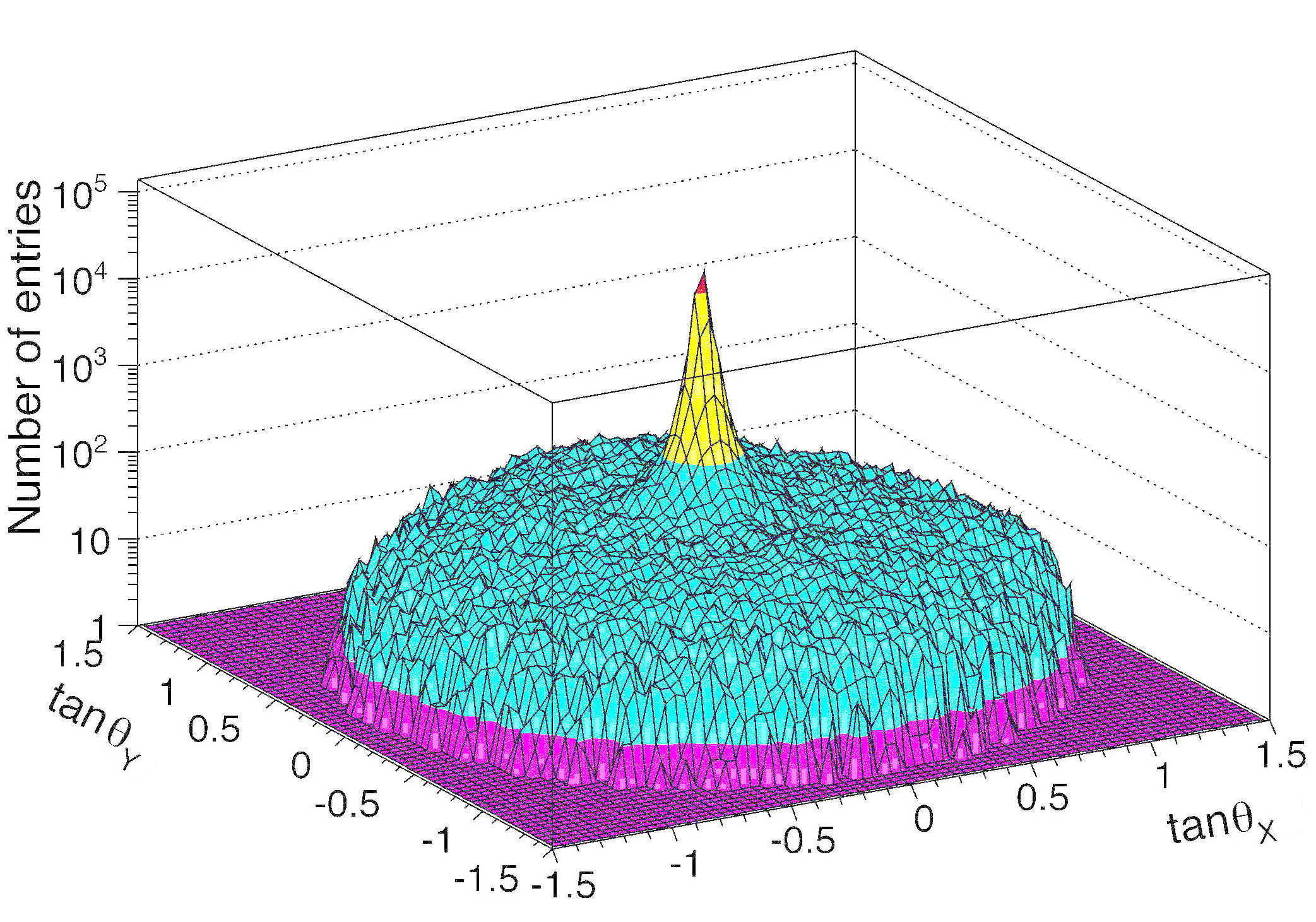}
}\centering}\hfill
\parbox{60mm}{\mbox{
\includegraphics[width=60mm]{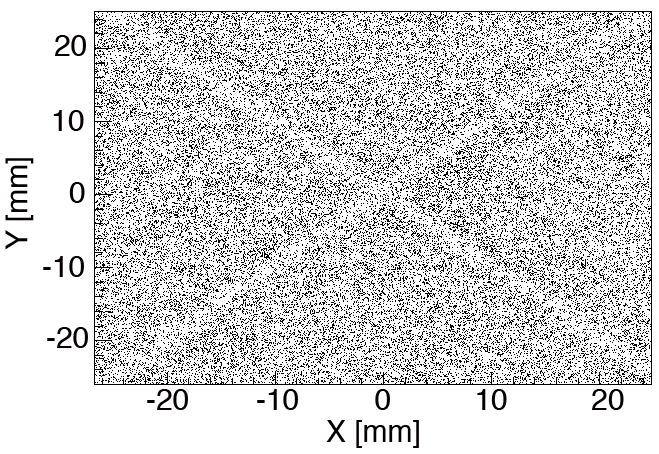}
}\centering} \caption[]{\it Left:  angular distribution of annihilation tracks. Plotted are the tangents of the projected track angles. Right:  spatial distribution of large angle tracks in an area of 5 $\times$ 5 cm$^2$ around the center.  \label{Spatial}}
\end{figure}

Figure~\ref{Spatial} (left) shows a typical two-dimensional angular distribution of the tracks. Plotted  in a logarithmic scale are the distributions of tan$\theta_x$ and tan$\theta_y$ of the track angles projected into the $x-z$ and $y-z$ planes. The high intensity at normal incidence ($\theta_x \sim\theta_y \sim 0$) is due to pions generated in the far upstream part of the beamline, while the distribution at large angles stems from annihilations in the vicinity of the emulsion detector. Figure~\ref{Spatial} (right) shows the spatial distribution of tracks with incident angles larger than 25$^\circ$. Scattering and absorption in the emulsion holder (bright cross) are clearly observed. The typical antiproton flux reaching the detector was 2000 cm$^{-2}$ per antiproton beam extraction.

\begin{figure}[htb]
\parbox{75mm}{\mbox{
\includegraphics[width=65mm]{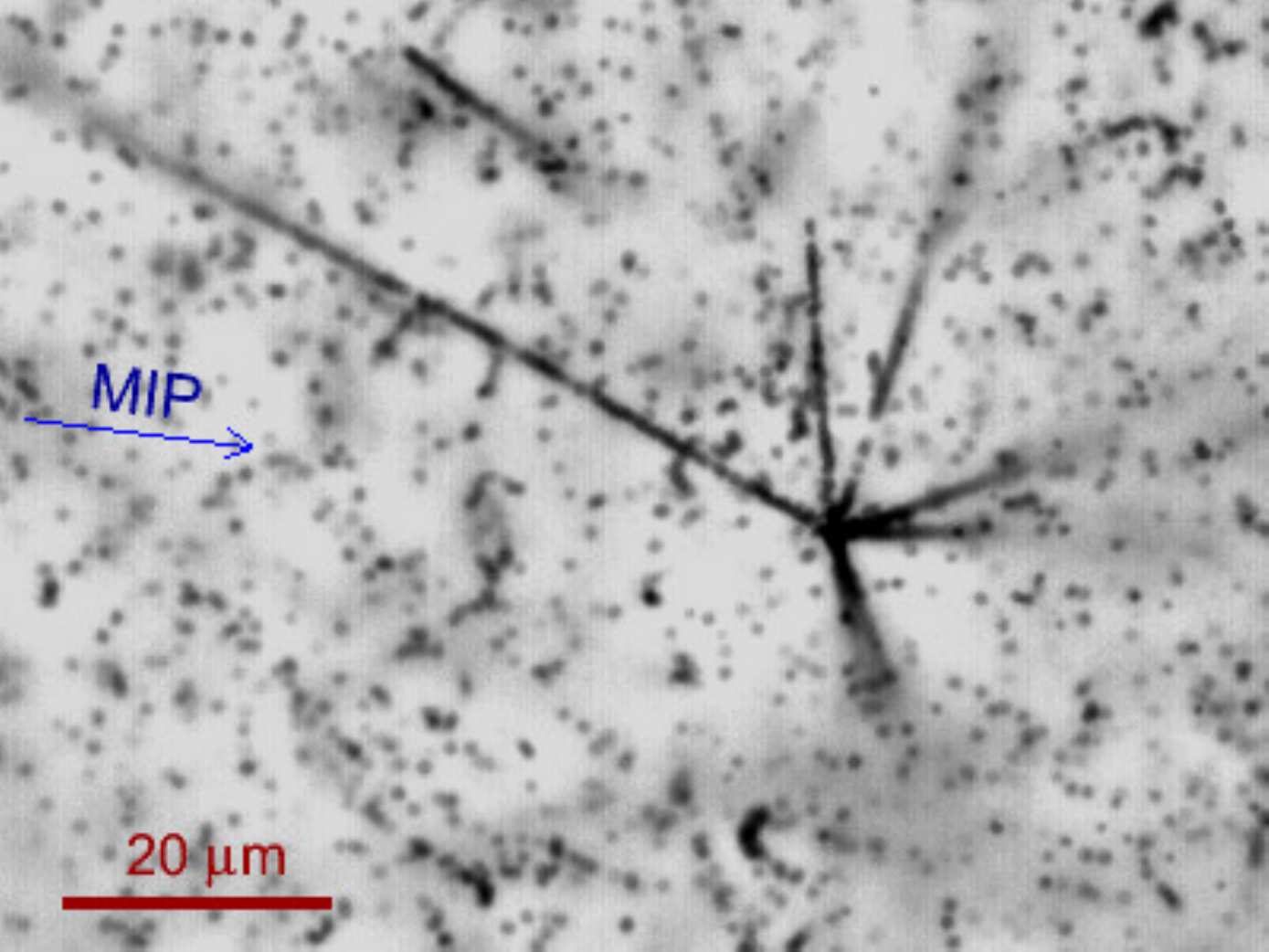}
}\centering}\hfill
\parbox{75mm}{\mbox{
\includegraphics[width=65mm]{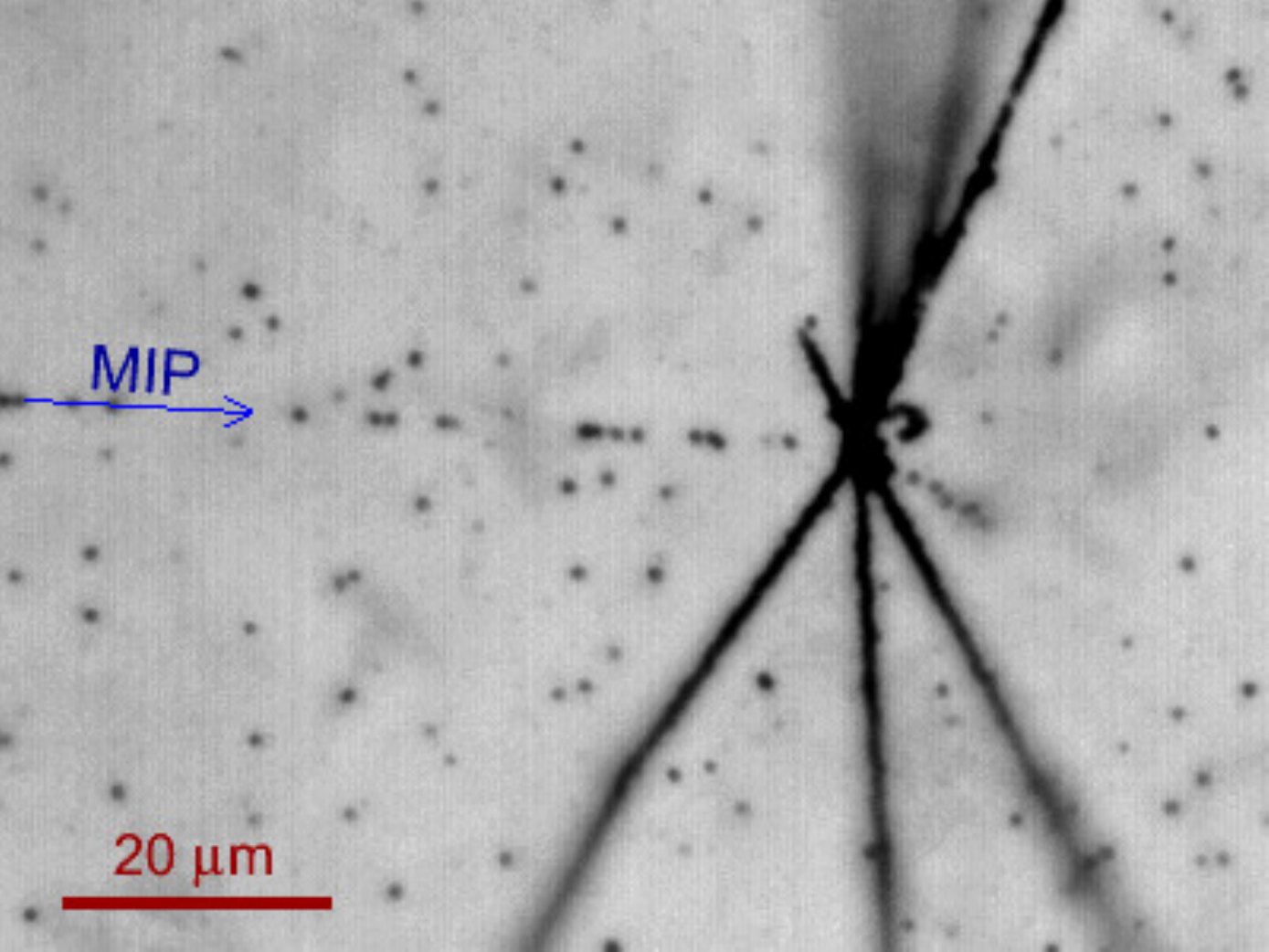}
}\centering} \caption[]{\it Typical tracks from minimum ionizing particles and from fragments emerging from the annihilation vertex, observed in the standard gel (left) and in the new gel (right). 
 \label{Events}}
\end{figure}

Typical annihilation vertices and tracks in the new gel are compared to those in the standard gel in figure~\ref{Events}. The standard gel  (left) was produced   in 2005. This explains the  high background accumulated since then. The new gel (right) was prepared in December 2012 at LHEP from a gel developed by the University of Nagoya. The heavily ionizing particles emitted from the annihilation vertices, mainly protons and heavy nuclear fragments, are clearly visible in both films. The faint tracks from MIPs (indicated by arrows) are presumably pions, barely seen in the standard gel while clearly observed in the new one, thanks to the higher signal-to-noise ratio.

\subsection{Annihilation on metal foils}
Next we  carried out measurements with a series of thin foils of varying compositions to (i) determine the achievable vertex resolution in the $\bar{g}$ experiment  and (ii) to measure the relative contributions from heavily ionizing particles (such as protons, deuterons, $\alpha$-particles
 and nuclear fragments) and MIPs, as a function of atomic number. Apart from obvious interests in nuclear physics this will provide a useful input to Monte Carlo simulations (such as GEANT4) and is also crucial to predict the reconstruction efficiency of our emulsion detector in the $\bar{g}$ measurement, as will be explained below. 

\begin{figure}[htb]
\parbox{75mm}{\mbox{
\includegraphics[width=50mm]{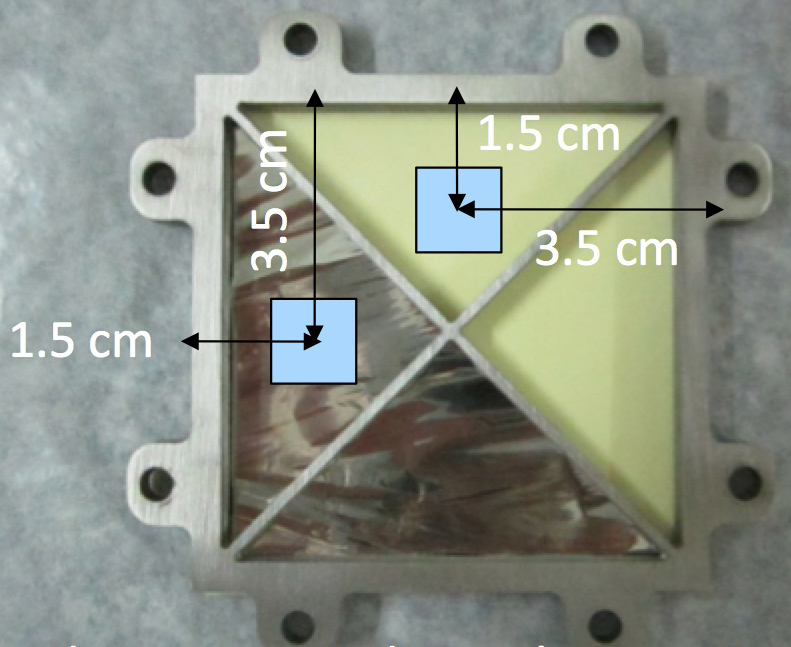}
}\centering}\hfill
\parbox{75mm}{\mbox{
\includegraphics[width=40mm]{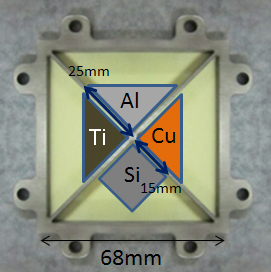}
}\centering} \caption[]{\it Left: setup used to measure the vertex resolution. Half of the surface is covered by a 20 $\mu$m thin stainless steel foil. The blue squares indicate the scanning area used for the present analysis. Right: arrangement of foils to determine the annihilation multiplicities in various metals. \label{Setup}}
\end{figure}

\begin{figure}[htb]
\parbox{75mm}{\mbox{
\includegraphics[width=75mm]{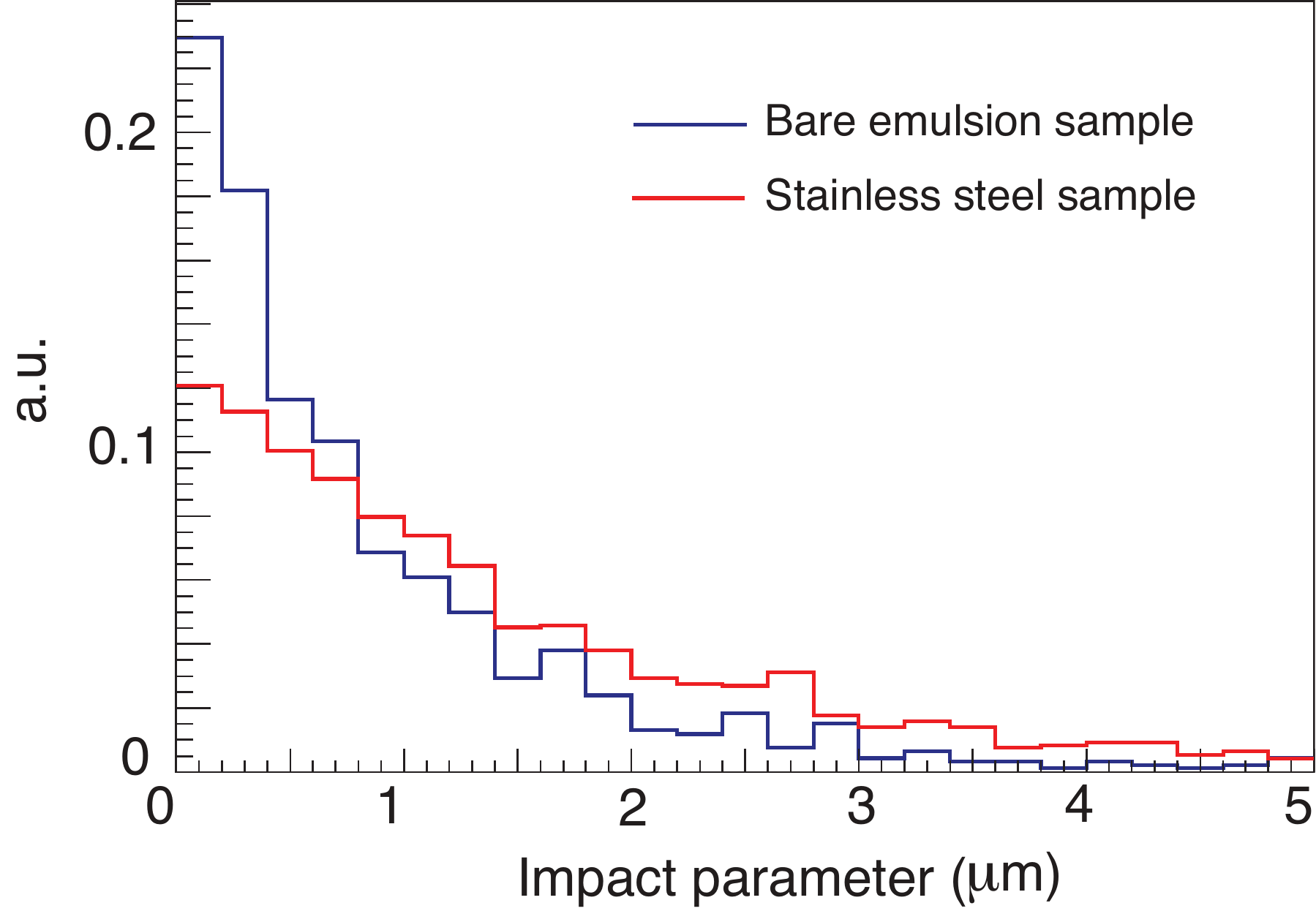}
}\centering}\hfill
\parbox{75mm}{\mbox{
\includegraphics[width=75mm]{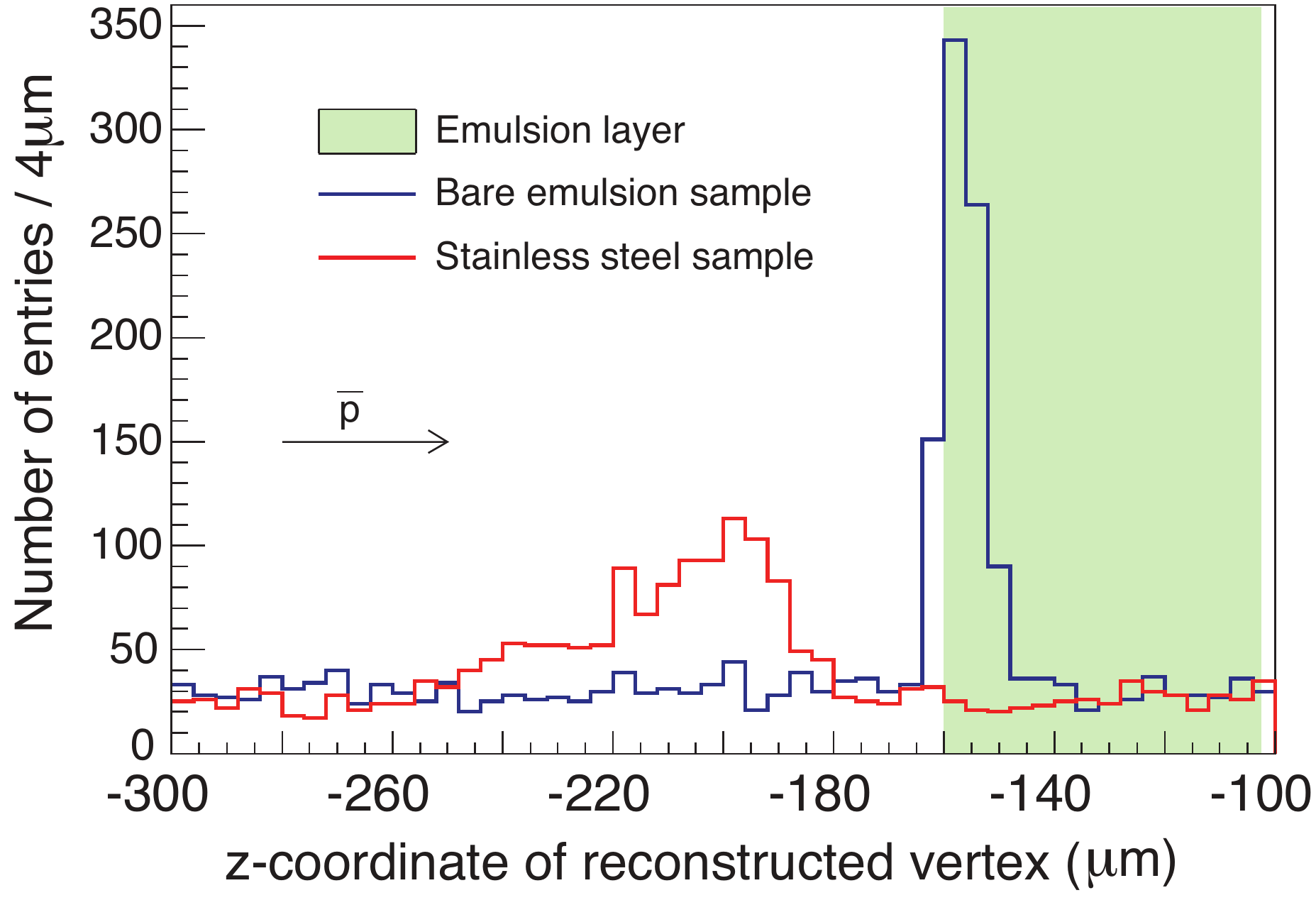}
}\centering} \caption[]{\it Left: impact parameter distribution for annihilation on the bare emulsion surface (standard gel) and in the stainless steel foil. The histograms are normalized to the same area.  A resolution of $\sim$1 $\mu$m can be achieved for the $\bar{g}$-measurement. Right: vertex resolution in the $z$-direction (perpendicular to the surface). \label{Resolutions}}
\end{figure}

The first tests were performed with standard gels, half of the emulsion surface being covered by a 20 $\mu$m thick stainless steel foil, while 
direct annihilation on the emulsion surface could be investigated from the other half (figure~\ref{Setup}, left). The vertex position was calculated by finding the coordinates of the point which minimizes the quadratic sum of the distances to all tracks. Figure~\ref{Resolutions} (left) shows the distribution of the distance between this point and  all tracks (impact parameter), which is a measure of the resolution with which the annihilation point will be determined in the $\bar{g}$-measurement. The figure~shows  that in principle (without systematic errors, such as film distortions, misalignments), an r.m.s resolution of $\sigma \simeq$ 1 $\mu$m on the vertical position of the annihilation vertex can be achieved with  a 20 $\mu$m steel window. 

\begin{figure}[htb]
\parbox{150mm}{\mbox{
\includegraphics[width=150mm]{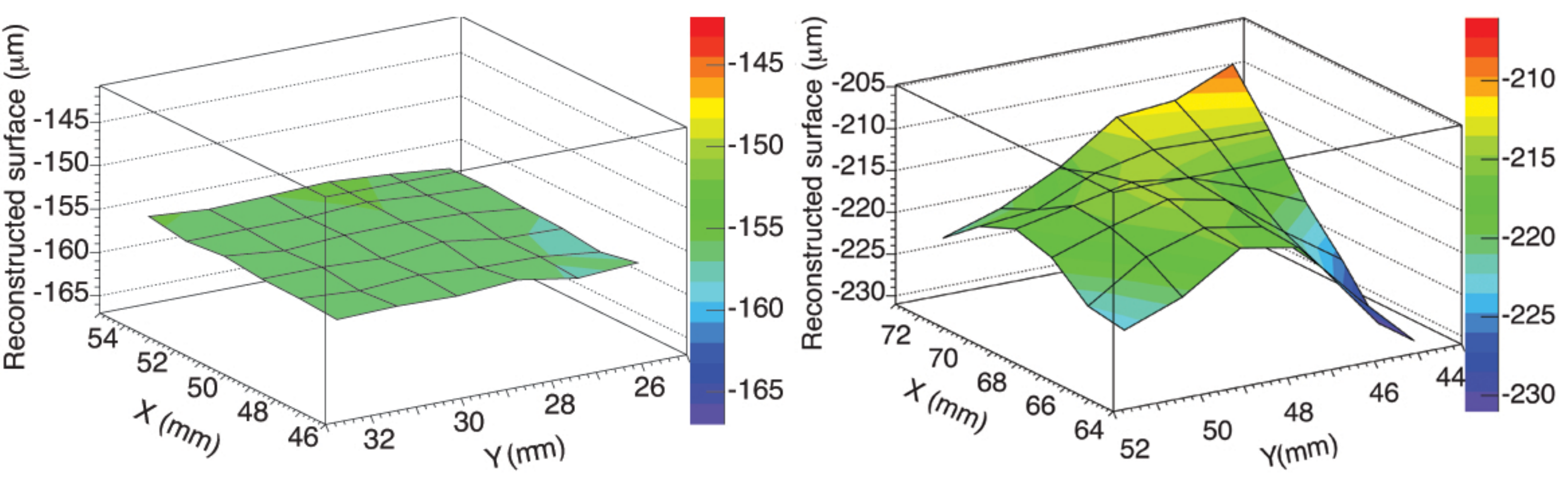}
}\centering}\hfill
\caption[]{\it  Topography of the emulsion surface (left) and of the stainless steel foil (right). The vertical scale refers to the distance in $\mu$m from the vertex to the center of the base. \label{Notflat}}
\end{figure}

Figure~\ref{Resolutions} (right) shows the distribution of the vertex position perpendicular to the film surface ($z$-direction). The flat distribution in figure~\ref{Resolutions} (right) is due to combinatorial background in the vertex reconstruction. The distribution has a sharp peak for annihilation in the emulsion, while the stainless steel sample shows a broader peak. This is due to the fact that the foil was not flat nor in perfect contact with the film surface. In fact, the distribution of the vertex position in $z$ is a measure of the surface flatness. Figure~\ref{Notflat} shows  the topography of the emulsion surface (left) and of the surface of the stainless steel foil (right) obtained from reconstructed annihilation vertices.   

\begin{figure}[htb]
\parbox{150mm}{\mbox{
\includegraphics[width=120mm]{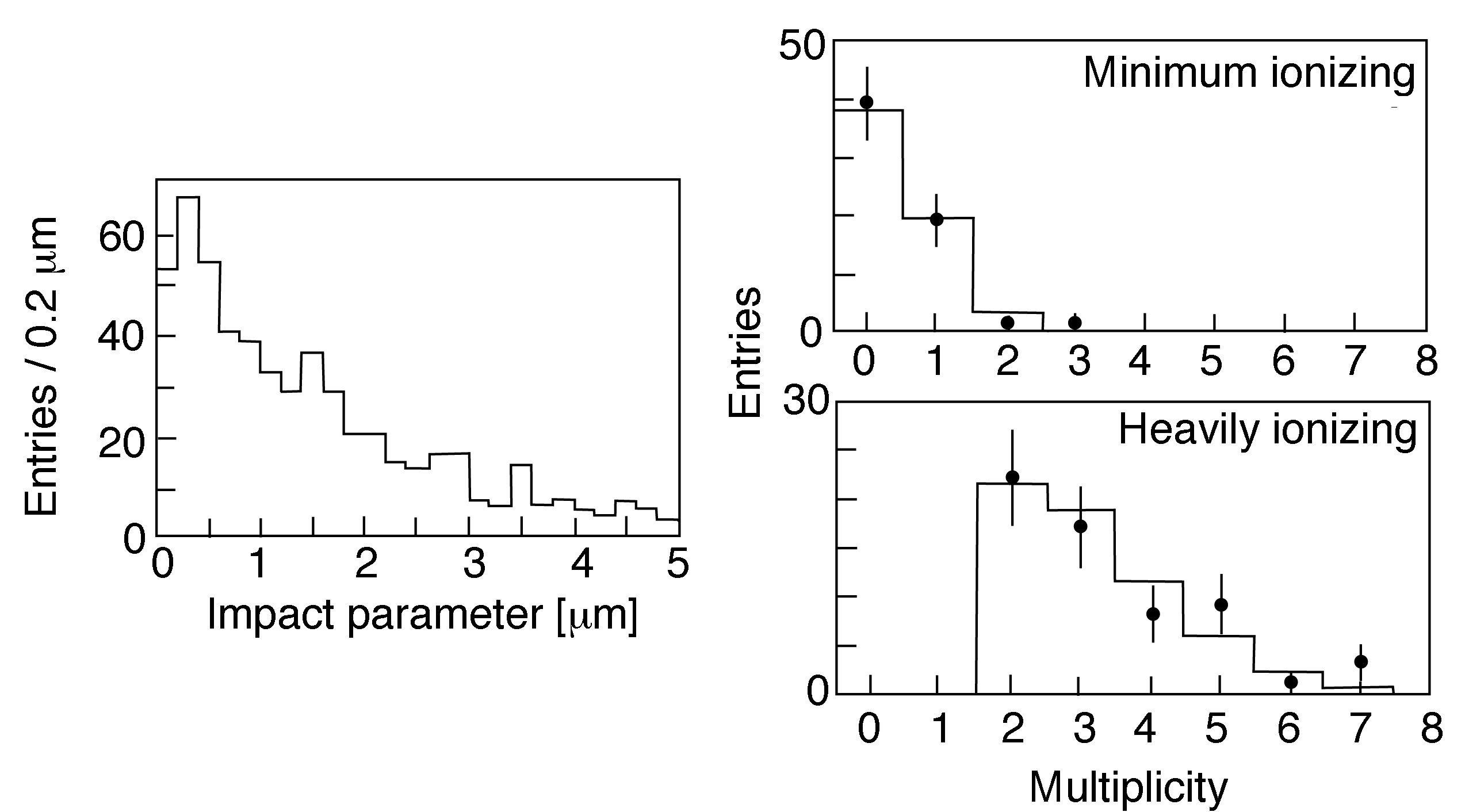}
}\centering}\hfill
\caption[]{\it Measured impact parameter distribution for a 6 $\mu$m aluminium foil (left) and multiplicity distribution for 60 events annihilating on the surface of the foil, for minimum ionizing and for heavily ionizing particles. The histogram is the Monte Carlo prediction from CHIPS.   
\label{Alplot}}
\end{figure}

Further tests were also performed with the standard gel and various thin foils made of Al, Si, Ti, Cu, Ag , Au and Pb. The foil arrangement for the Al, Cu, Si, and Ti measurements is shown in figure~\ref{Setup} (right). Here we report only on the result  for 6 $\mu$m aluminium (99\% purity), the analysis of the other foils being  the subject of a forthcoming paper. The aluminium foil was in direct contact with the detector consisting of five stacked emulsion detectors. We required at least two tracks from heavily ionizing particles (mostly protons and also nuclear fragments). These tracks are characterized by a continuous black line of clusters, while MIPs lead to  aligned but separated grains and hence to faint tracks  (see figure~\ref{Events}). 
An annihilation vertex in the foil was then predicted from the distance of closest approach. A scan for further tracks (including minimum ionizing ones) emerging from the putative vertex was then performed, followed by a visual inspection to reject faked tracks.  Figure~\ref{Alplot} (left) shows the impact parameter distribution. A similar resolution as with the stainless foil of figure~\ref{Resolutions} is obtained, $\sigma$ $\simeq$1.2 $\mu$m. 
Figure~\ref{Alplot} (right) shows the multiplicity distributions for  minimum ionizing and for heavily  ionizing particles. The histograms are the Monte Carlo predictions by the CHIPS package embedded in GEANT4 \cite{CHIPSref}. Heavily ionizing particles are selected by requiring  an energy loss of more than 3 MeV$\cdot$g$\cdot $cm$^{-2}$. The data for aluminium and the simulation are in good agreement.

\subsection{Annihilations on the emulsion surface}
Direct annihilations on the emulsion surface were also studied with the new gel.  One of the microscopes at the LHEP was devoted to AEgIS data analysis. The microscope was equipped with an $XYZ$ computer driven high precision stage, a 1.3 M-pixels high speed camera and an oil objective lens with a magnification of 50, leading to views  of 303 $\times$ 242 $\mu$m$^2$, or to a pixel pitch of 0.2365 $\mu$m. The regions of interest on the emulsion surface were fully digitized by the microscope. Samples of 19'625 views/film were taken, covering surfaces of 3 $\times$ 3 cm$^2$, and producing files  of 1.1 TB size. The digitizing time was 10 hours, mainly  limited by disk access time.
The digitized data were displayed on the PC monitor by an event display (the so-called ``Virtual Microscope'') which can  deal with TB data sizes  through  dynamic data access with multithread programming. The image can be inspected and controlled as if the user were looking directly into the microscope. 

\begin{figure}[htb] 
\parbox{150mm}{\mbox{
\includegraphics[width=80mm]{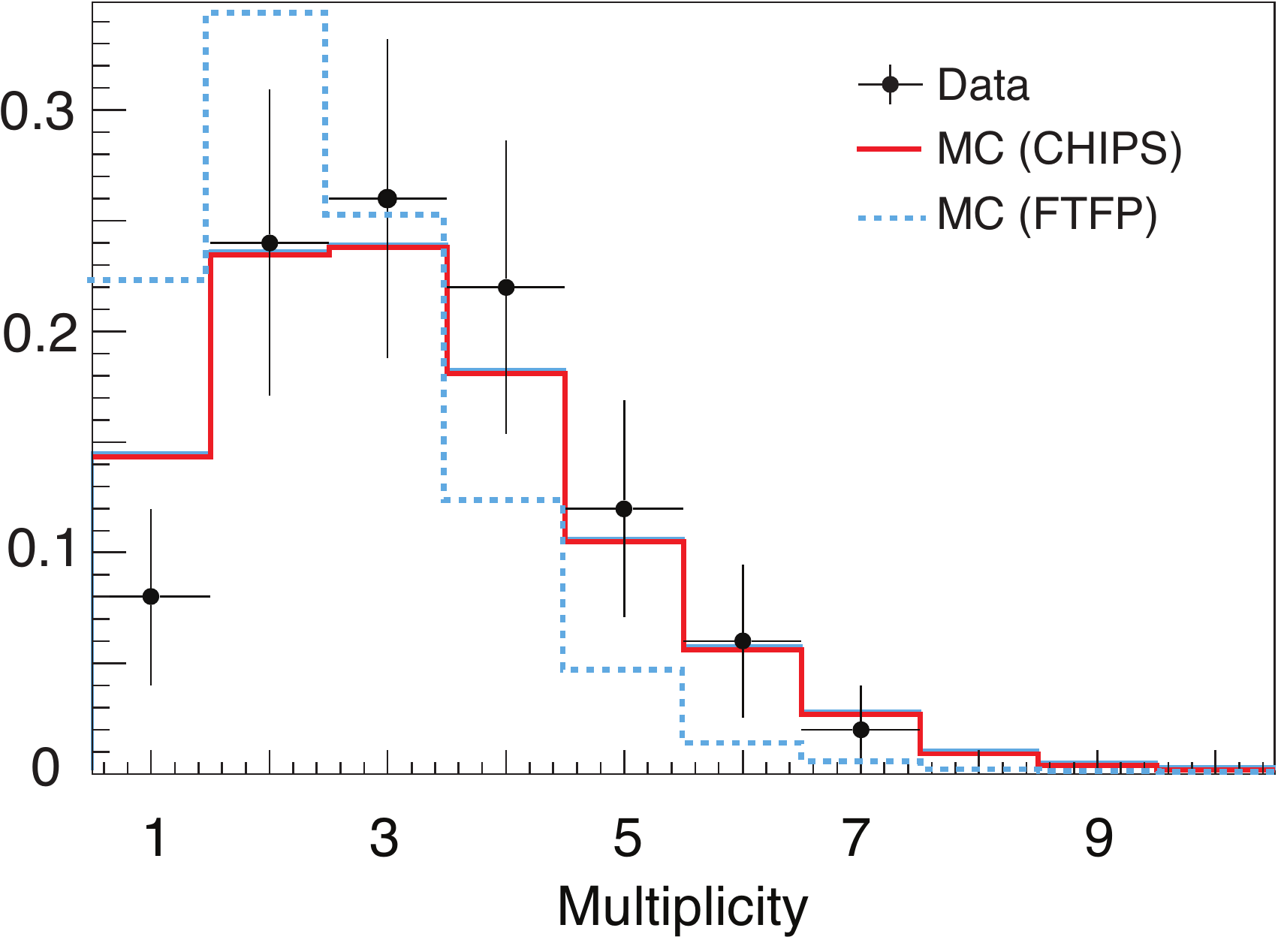}
}\centering}\hfill
\caption[]{\it Multiplicity distribution of ejectiles emitted isotropically into a solid angle $2\pi$ in antiproton annihilation at rest in emulsions (new gel), for data and for Monte Carlo predictions based on CHIPS (red continuous histogram)  and FTFP (blue dashed  histogram). The distributions are normalized to one.
\label{CHIPSTotal}}
\end{figure}

\begin{figure}[htb]
\parbox{150mm}{\mbox{
\includegraphics[width=80mm]{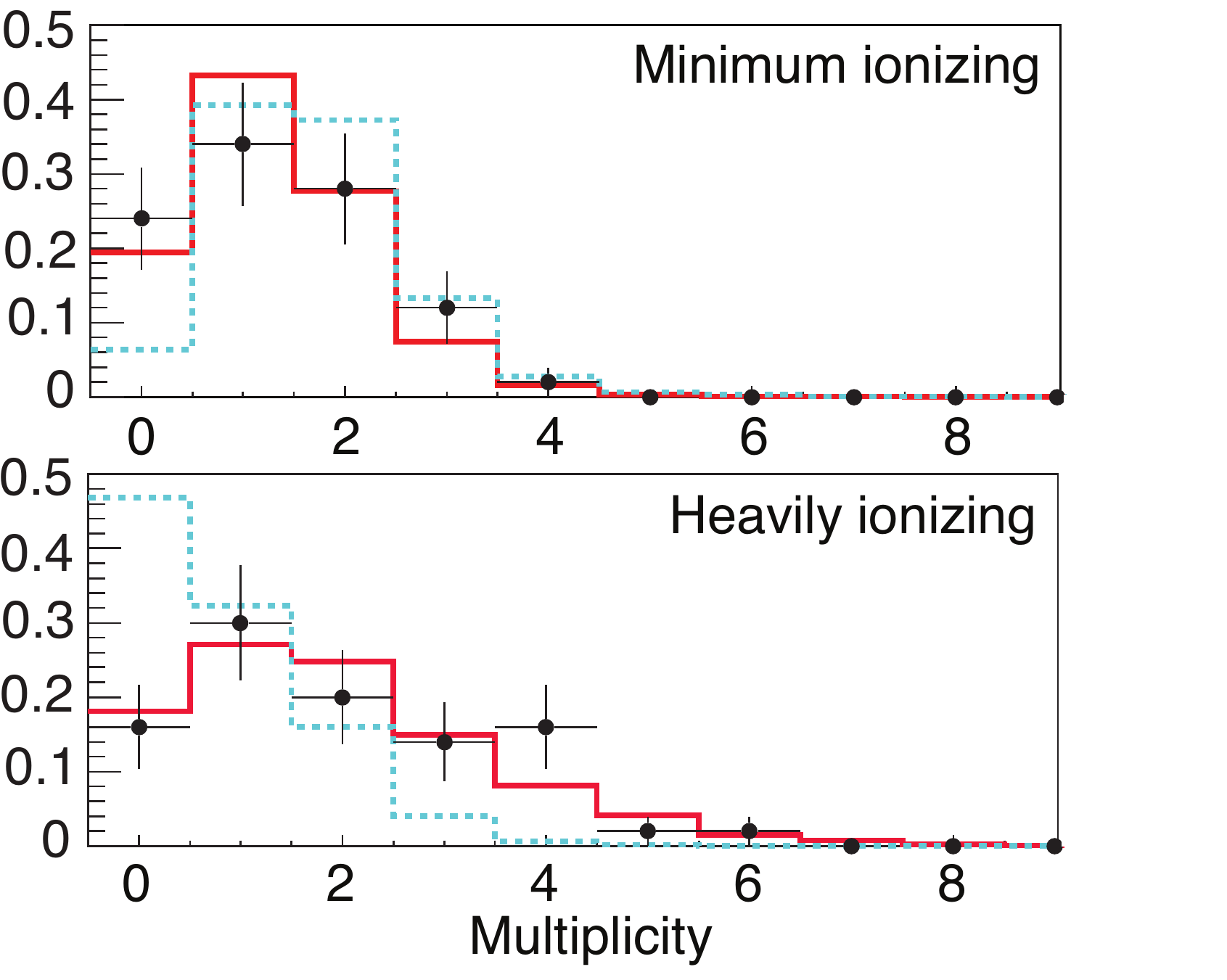}
}\centering}\hfill
\caption[]{\it Multiplicity distributions as for figure~\ref{CHIPSTotal}, but separately for MIPs (top) and for heavily ionizing ejectiles (bottom). The red continuous histograms show Monte Carlo predictions from  CHIPS, the blue dashed histograms predictions from FTFP. The distributions are normalized to one.
\label{CHIPSMips}}
\end{figure}

The multiplicity distribution for annihilation on the emulsion surface was studied with the new gel and compared to Monte Carlo simulation. The selection was performed as follows: Events were required to have at least one track penetrating a depth of more than 5 $\mu$m  from the emulsion surface (``long'' track) and at least a second one traversing more than 2 $\mu$m of emulsion material (``short'' track). The selection of short tracks ensured a coverage of nearly 2$\pi$ solid angle for annihilation particles emerging from the surface of the emulsion. The selection was performed manually, thus reaching a track recognition efficiency of nearly 100\%. A vertex was then identified with an efficiency of 96\%. The multiplicity distribution of the long tracks is shown in figure~\ref{CHIPSTotal} and compared to predictions by CHIPS \cite{CHIPSref}. Very good agreement between data and simulation is observed. The measured average multiplicity is 3.3 $\pm$ 0.2. Figure~\ref{CHIPSMips} shows the multiplicity distributions separately for MIPs (such as pions and electrons) and for heavily ionizing particles (such as protons, deuterons, and $\alpha$-particles). Particles with energy losses below 3 MeV$\cdot$g$\cdot$cm$^{-2}$ were defined as being minimum ionizing. Again, excellent agreement between data and GEANT4 (CHIPS) is observed. The measured average multiplicity is 1.3 $\pm$ 0.1 and 2.0 $\pm$ 0.2, respectively. A systematic error of 0.1 is ascribed to take into account uncertainties in assigning tracks to either class of ionizing particles.

Figs. \ref{CHIPSTotal} and \ref{CHIPSMips} also show the predictions from  Monte Carlo simulation based on FTFP (Fritiof) \cite{FTFPref}  (dashed lines), as an alternative to CHIPS. Clearly CHIPS performs much better than FTFP in describing antiproton annihilation at rest. The latter is adequate to describe annihilations at energies above 100 MeV/c \cite{Galoyan}. We should, however, warn the reader that more positive than negative pions are produced by CHIPS, which is surprising since  (negatively charged) antiprotons can annihilate on neutrons. Indeed, experimental data show that more negative than positive pions are produced in low energy antiproton annihilations \cite{Bendi}.

\section{Monte Carlo simulations}
\subsection{Resolution and efficiency of the emulsion detector}
As shown above, our multiplicity results for minimum  and heavily ionizing particles are in very good agreement with the CHIPS package. We have therefore simulated the expected performance of the emulsion detector in the $\bar{g}$ measurement using  GEANT4 (CHIPS) to model annihilation on nuclei and the nuclear de-excitation. The ingredients in the simulation are as follows: We choose antiprotons of 1 eV (the minimum possible energy in the package) impinging randomly on a 50 $\mu$m silicon or, alternatively, on a 5 $\mu$m thick titanium foil.  A single emulsion detector is used (standard gel, 44 $\mu$m thick, double sided on a glass or plastic base). A smearing of $\pm$3 $\mu$m is introduced to model the intrinsic track resolution in the $z$-direction. The intrinsic resolution perpendicular to the track is around 0.05 $\mu$m, hence negligible.  

\begin{figure}[htb]
\parbox{150mm}{\mbox{
\includegraphics[width=110mm]{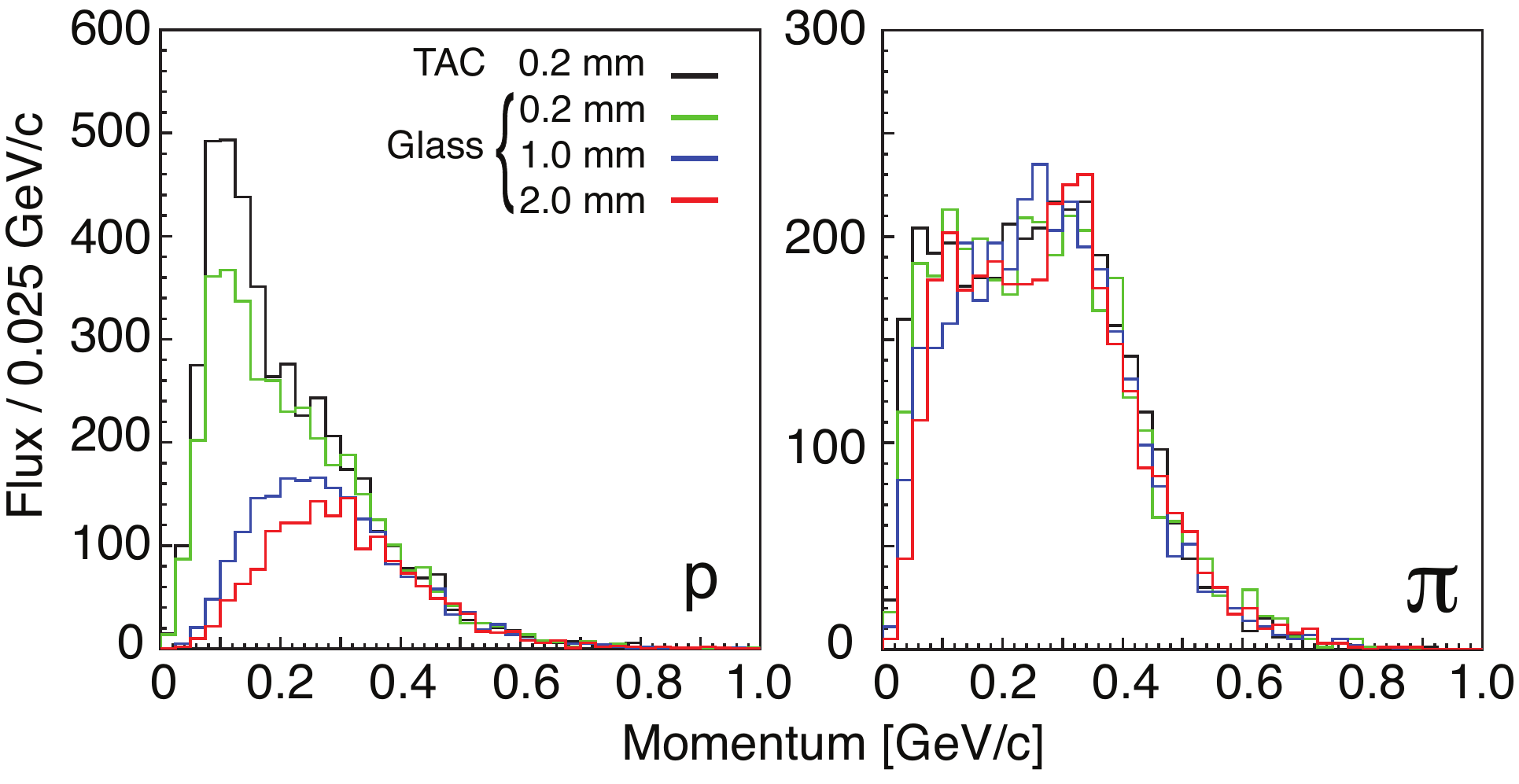}
}\centering}\hfill
\caption[]{\it  Momentum distribution of basetracks for protons (left), respectively pions (right), for various glass thicknesses and for a plastic (TAC) base.
\label{Gapsim}}
\end{figure}

A  gap is assumed between the foil and  the detector to prevent  mechanical contact and to allow the possible insertion of very thin superinsulation foils to thermally shield the emulsions from the low temperature region. The glass base has to be thin enough to enable the observation of at least two microtracks.  Figure~\ref{Gapsim}  shows the momentum distribution of the protons and of the pions for various glass thicknesses. The efficiency to detect protons decreases with increasing glass thickness, as expected, while pions are not affected. This leads to a decrease of detection efficiency, which can  be compensated by increasing the acceptance of the microscope. 

\begin{figure}[htb]
\parbox{150mm}{\mbox{
\includegraphics[width=70mm]{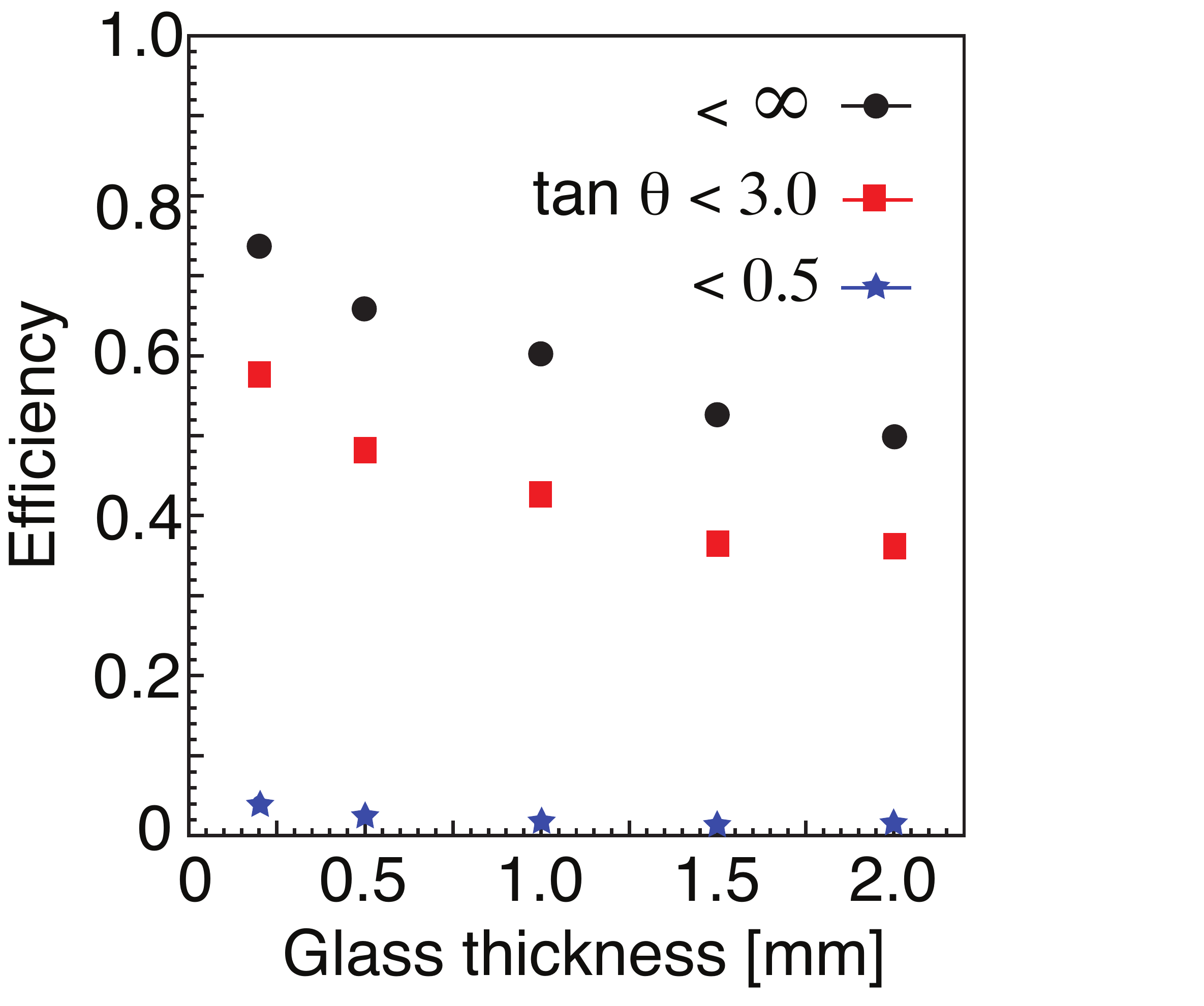}
}\centering}\hfill
\caption[]{\it  Efficiency to detect a vertex as a function of the glass thickness for various maximum incident angles $\theta$.  \label{AngularAcceptance}}
\end{figure}

Figure~\ref{AngularAcceptance} shows the vertex finding efficiency when requiring at least two basetracks, for different angular acceptances. The angular acceptance $\tan\theta < 0.5$  used in the analysis of neutrino data \cite{OPERA} is shown for comparison in figure~\ref{AngularAcceptance}. High energy neutrinos  hit the emulsion detector perpendicularly, and daughter particles produced by the interaction are forward boosted. Both scanning procedure and reconstruction software were designed to maximize the efficiency at small angles. In contrast, the angular distribution of the annihilation products is isotropic in AEgIS, and  we had therefore to modify the scanning procedure and the reconstruction software to analyze the data presented in this article. 

The uncertainty in the vertex resolution along the direction of the gravitational force and the width of the vertex distribution along $z$ increase with gap width,  due to multiple scattering in the foil and to the lever arm effect. The loss of resolution along $z$  leads to a decrease in the efficiency to reconstruct a vertex compatible with annihilation on the surface of the foil. Hence the foil must be kept as thin as possible and the gap reduced to the minimum technically feasible.

\begin{table}[htb]
\begin{center}
\begin{tabular}{r  r r r r}
\hline
& \multicolumn{2}{c}{Si} & \multicolumn{2}{c}{Ti} \\
\hline
Gap &  $\sigma$ & $\epsilon$ & $\sigma$ & $\epsilon$ \\
 $ [\mu\textrm{m}]$ & $[\mu\textrm{m}]$  & [\%] &$ [\mu\textrm{m}]$  & [\%]  \\
\hline
0 & 1.0& 53 & 0.8 &45 \\
100 & 1.4 & 49&1.3 & 44\\
250 & 2.2 & 46 &1.9 &41\\
500 &  2.9& 41 &2.8&37 \\
1000 & 4.2 &37 & 4.2 & 33\\
\hline
0 & 1.1&46 &0.8 & 41\\
100  & 1.5&44 &1.2 & 39\\
250 & 2.1 &41 & 1.6& 35\\
500 & 2.5 &37 &2.2 & 33\\
1000 &3.7 &34 & 2.9&30\\
\hline
\end{tabular}
\caption[]{\it Resolutions $\sigma$ and vertex finding efficiency  $\epsilon$ on 50 $\mu$m thick silicon and 5 $\mu$m thick titanium foils. The top half of the table is for a 500 $\mu$m glass base, the bottom half for a 1 mm glass base.  }
\label{table:gapglass}
\end{center}
\end{table}

The results for the vertex resolution $\sigma$ (assuming gaussian distributions) and the efficiency $\epsilon$ are summarized in table \ref{table:gapglass} for  500 $\mu$m and 1 mm glass bases, and for 50 $\mu$m thick silicon and 5 $\mu$m thick titanium foils. Required are at least two tracks traversing the glass base (basetracks). We assume for the angular acceptance a realistic value of $\tan\theta$ < 3. We require the vertex coordinate to be compatible with the surface of the foil:  Since the width of the vertex distribution along $z$ increases with gap width, we apply a variable cut in $z$ of 10 $\mu$m + 0.01 $\times$ gap [$\mu$m] between the reconstructed vertex and the surface of the foil. Note that the resolutions for a 1 mm glass base can be better than for 500 $\mu$m. This is due to the larger gap between the two microtracks (distance between points A and B in figure~\ref{Glossary}). 
For the base a  thickness of 1 mm could be chosen, below which glass plates would be too fragile. For example, a gap width of 250 $\mu$m leads to a typical r.m.s. resolution of 2 $\mu$m for a vertex reconstruction efficiency of about 40\%. Higher vertex reconstruction efficiencies (up to $\sim$70\%) are possible when relaxing the $z$-cut, however at the expense of worsening the resolution. Optimizations of the vertex reconstruction algorithms  are underway. 

\subsection{Performance of the emulsion detector for the $\bar{g}$ measurement}
\label{sec:42}
The influence of the spatial resolution of the $\overline{\textrm{H}}$ detector on the measurement of $\bar{g}$ has been evaluated by Monte Carlo simulation. The $\overline{\textrm{H}}$ atoms were uniformly generated in a cylinder of 3 mm transverse radius and 2 cm axial length, with initial velocities distributed according to the Maxwell-Boltzman distribution corresponding to a temperature of 100 mK. The $\overline{\textrm{H}}$ atoms were then submitted to a constant acceleration. For the moir\'e deflectometer we assumed a distance $L$ of 50 cm (figure~\ref{AEGISPrinz}) and a pitch $p$ = 40 $\mu$m between horizontal slits of 12 $\mu$m width \cite{Oberthaler}. The first grating was placed at a distance of 1 m from the source.  A Gaussian smearing was then applied to the annihilation vertex to simulate the detector resolution.

\begin{figure}[htb]
\parbox{150mm}{\mbox{
\includegraphics[width=80mm]{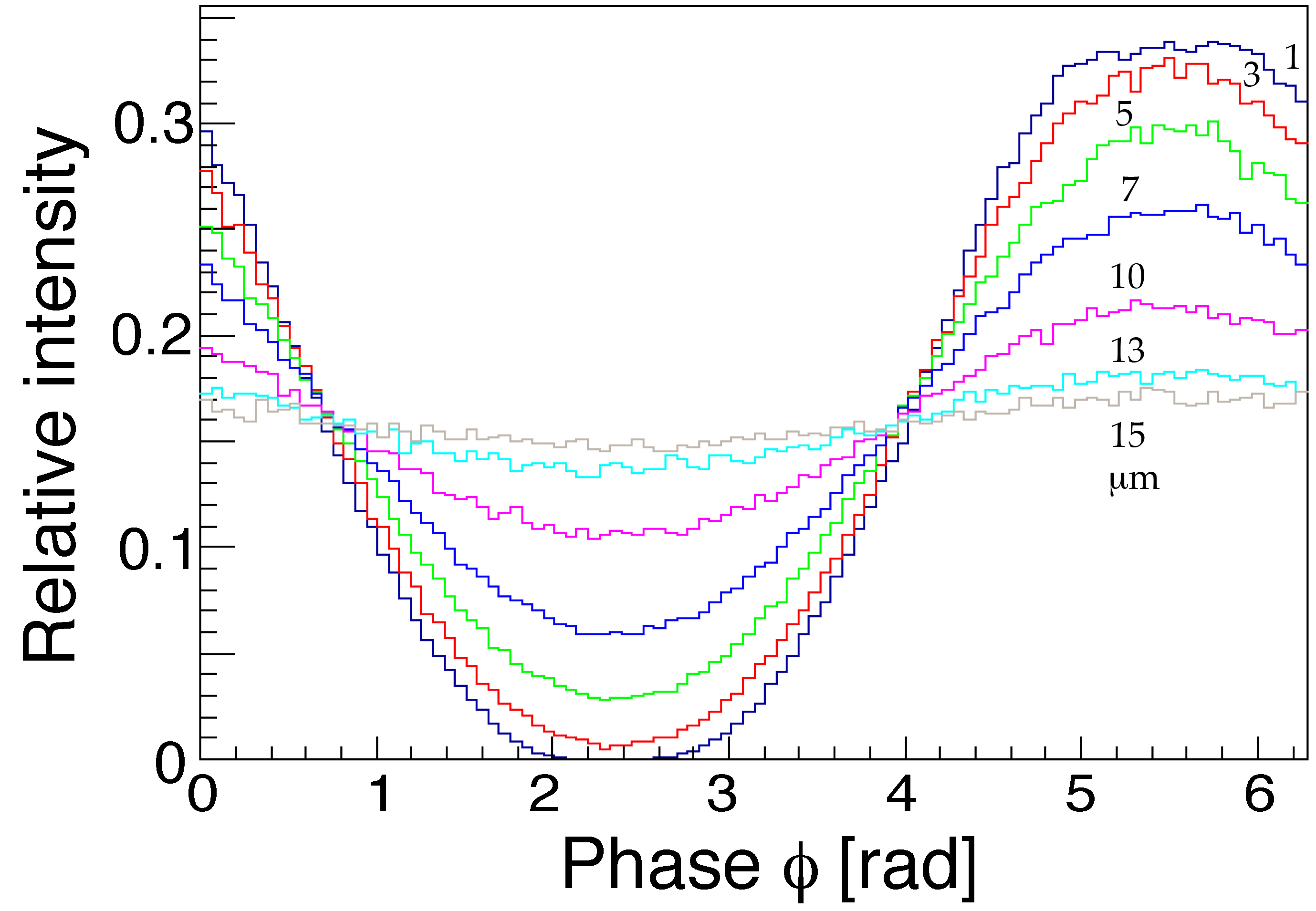}
}\centering}\hfill
\caption[]{\it  Simulated intensity distribution of reconstructed vertices for various vertex resolutions between 1 and 15 $\mu$m. 
\label{Contrast}}
\end{figure}

The unbinned maximum likelihood method was used to evaluate the uncertainty on $\Delta g/g$ as a function of the number of annihilations, for various vertex resolutions. Assuming for $g$ (= $\bar{g}$) a  value of 9.81 m/s$^2$ we repeated the experiment 300 times. To fit the vertical vertex coordinate we used the following probability density function for each annihilation with vertical vertex coordinate $x_i$:

\begin{equation}
P(x_i) = a_0(T)+\sum_{j=1}^5 a_j(T)\cos\left(j\left[\frac{2\pi}{p}x_i+\frac{2\pi}{p}gT^2_i\right]\right),
\end{equation}
where $T$ is the time of flight. Thus we described the intensity pattern with cos-functions up to the 5$^{\rm th}$ harmonic. The amplitudes $a_j(T)$ ($ j$ = 0,...,5) were obtained from the fit for fixed values of $T$   
and then parameterized as a function of $T$ with appropriate interpolating functions.  We show in figure~\ref{Contrast} the expected intensity distribution of the annihilation vertex for various spatial resolutions and, as an example, for a particle velocity of 700 m/s. Plotted is the phase $\phi = (2\pi/p) \Delta x$, where $\Delta x$ = $g T^2$ and $T$ is the time of flight. 

\begin{figure}[htb]
\parbox{75mm}{\mbox{
\includegraphics[width=75mm]{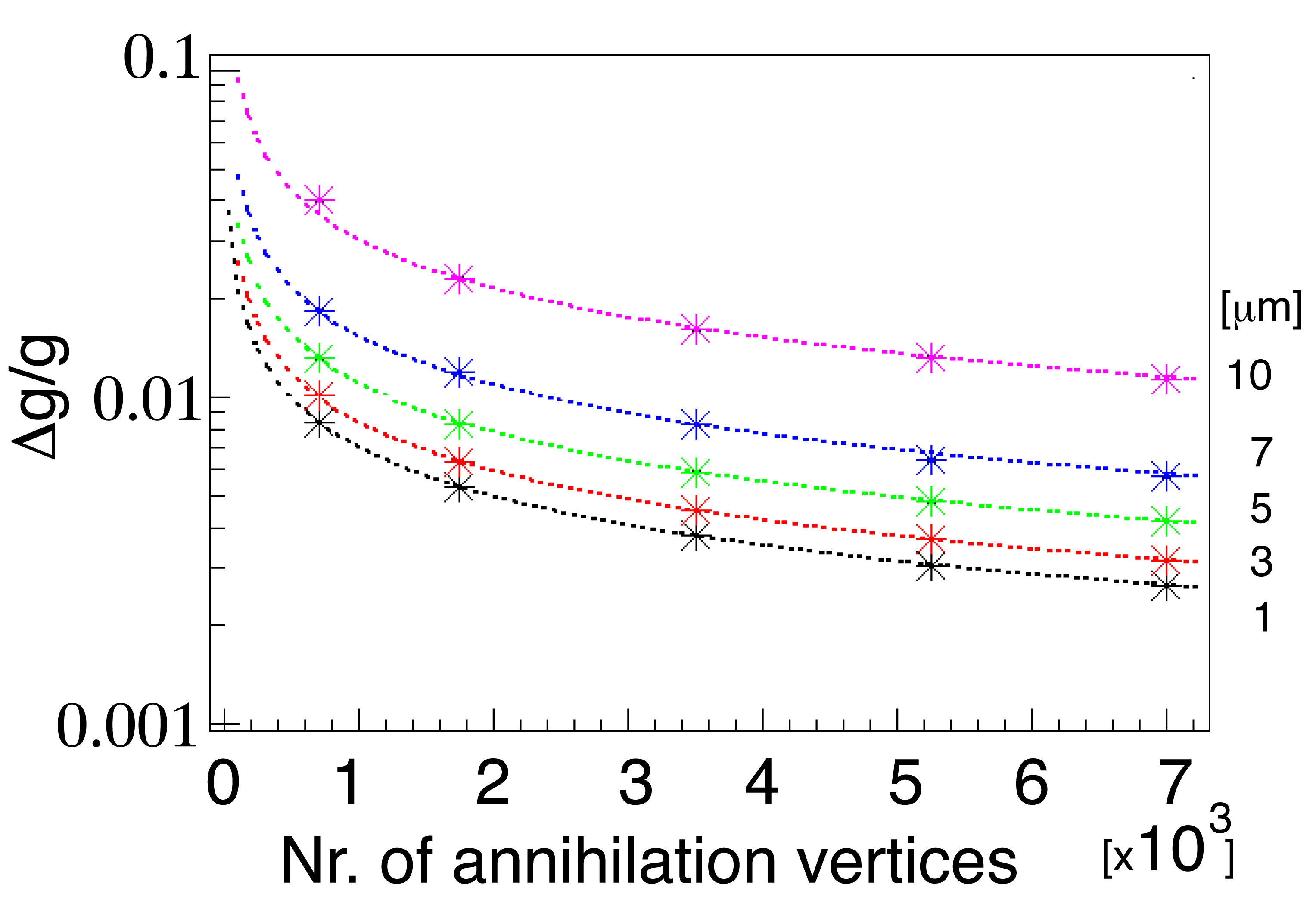}
}\centering}\hfill
\parbox{75mm}{\mbox{
\includegraphics[width=75mm]{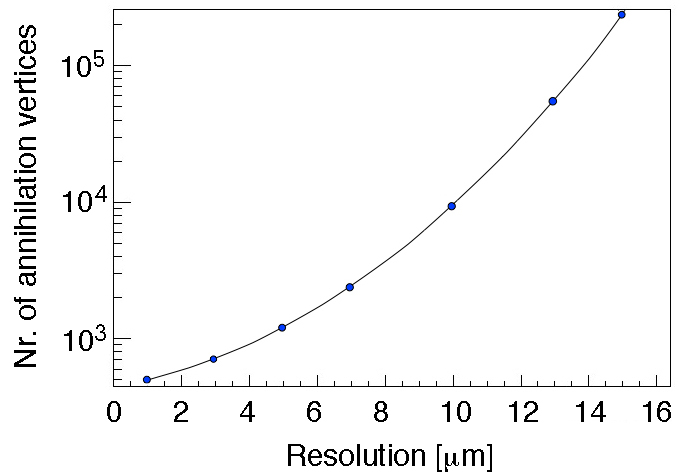}
}\centering} \caption[]{\it 
Left: relative precision on $g$ vs. number of annihilation vertices for various vertex resolutions from 1 to 10 $\mu$m.   Right: Number of annihilation vertices required to reach 1\% on $\Delta g/g$, vs. vertex resolution (the curve is an eyeball fit to the points). \label{Resdeltag}}
\end{figure}

The required number  of reconstructed $\overline{\textrm{H}}$ annihilations on the foil in front of the emulsion detector to reach the precision $\Delta g/g$ is shown in figure~\ref{Resdeltag} (left) for various resolutions. For resolutions larger than the slit width the contrast decreases quickly and, accordingly, the size of the required data sample rises rapidly (figure~\ref{Resdeltag}, right).  For a 1\% relative error on $g$  one needs  with a resolution of 2 $\mu$m a sample of $\sim$600 reconstructed  and time-tagged annihilations. 
The improvement in reso\-lution compared to the original proposal \cite{propaegis} by a factor of 5 -- 10 means that the experiment can be performed in a much shorter time. Alternatively, other experimental constraints can be relaxed, such as the challenging 100 mK required to limit the transverse $\overline{\textrm{H}}$ momentum. 

\section{Outlook}
A proof of principle using a miniature moir\'e deflectometer ($L$ = 25 mm, pitch $p$ = 40 $\mu$m, slit width 12 $\mu$m) was also performed during the beam tests. Several gratings  were used, as well as gratings in contact with the emulsion detector. Results are the subject of a forthcoming publication \cite{mini}.

In future, the digitized data from emulsion scanning is to be processed by means of a General Purpose Computing on Graphic Processing Unit  (GPGPU), a technology available through NVIDIA CUDA \cite{cuda}. A processor  equipped with 2 GPU, each with 2'688 cores (NVIDIA, GeForce GTX TITAN) has being installed in our scanning laboratory.   A tracking algorithm for low energy particles from antiproton annihilation stars is  under development and improvements in the track finding efficiency (such as increasing the solid angle acceptance towards large angle tracks) are underway. For the $\bar{g}$ measurement we plan to use a detector with an emulsion surface of 20 $\times$ 20 cm$^2$. The automatic scanning and digitizing will take about three days to process with one microscope (while six microscopes are currently available at the LHEP).

The use of photomasks \cite{Kimura} to reduce contributions from systematic errors to the vertex resolution, due to film distortions at development time, have been described in an earlier paper \cite{Emul} and are presently under study.

To limit liquid helium consumption and thermally shield the $\overline{\textrm{H}}$ source we might have to operate the emulsions below room temperature, possibly even at  77 K. Corresponding R\&D tests are underway to determine with cosmic rays and electron sources the performance of our emulsions (in particular the detection efficiency) as a function of temperature. We note that emulsions have been operated in the past at temperatures as low as 24 K, e.g. for neutrino interactions in deuterium bubble chambers \cite{Smart}. Tests in liquid nitrogen and even liquid hydrogen have also been reported (for a review see \cite{EMgroup}, where further references can be found).
 
\section{Conclusions}
Emulsion detectors can measure the annihilation vertex of antihydrogen atoms in the AEgIS experiment with very high precision. We have presented here results for emulsion films operating in vacuum at the CERN low energy antiproton decelerator.  We have shown data using a promising new emulsion gel with typically a factor of two higher grain density than used before in  emulsion detectors. This increases the efficiency for the detection of minimum ionizing particles such as the pions emitted in antiproton annihilations.
We have compared our measured particle multiplicites from annihilations in emulsions and aluminium with the Monte Carlo packages CHIPS and FTFP in GEANT4. CHIPS provides a  better description of the data than FTFP. We have estimated the resolution and data taking time needed to achieve the AEgIS goals. We obtain vertex resolutions $\sigma$ in the range $\sim$1 -- 2 $\mu$m which, when combined with  time of flight measurements, lead to an order of magnitude reduction of the data taking time originally foreseen to reach the goal of 1\% uncertainty in $\Delta g/g$. 

\newpage
\section*{Acknowledgments}
We wish to warmly acknowledge our technical collaborators Luca Dassa, Roger H\"anni and Jacky Rochet. We also thank the University of Nagoya for providing the ingredients needed in the preparation of the new gel.


\begin{thebibliography}{9}
\bibitem{propaegis} G. Drobychev {\it et al.}, 
http://doc.cern.ch/archive/electronic/cern/preprints/ \\spsc/public/spsc-2007-017.pdf; \\A. Kellerbauer {\it et al.}, 
Nucl. Instr. and Meth.  {\bf B 266} (2008) 351
\bibitem{Adelberger} E. G. Adelberger {\it et al.}, Progress in Particle and Nuclear Physics {\bf 62} (2009) 102
\bibitem{Goldman} M. M. Nieto and T. Goldman, Phys. Rep. {\bf  205} (1991) 221
\bibitem{Fischler} M. Fischler, Joe Lykken, T. Roberts, report FERMILAB-FN-0822-CD-T (2008)
\bibitem{Adelantimatter} T. A. Wagner {\it et al.}, prep. arXiv: 1207.2442  (2012)
\bibitem{Alves} D. S. Alves, M. Jankowiak and P. Saraswat, prep. 0907.4110 (2009)
\bibitem{Karshen1} S. G. Karshenboim, prep. arXiv: 0811.1009 (2008)
\bibitem{Kostel} V. A. Kosteleck\'y and J. T. Tasson, Phys. Rev. {\bf D 83} (2011) 016013
\bibitem{Cassidy} P. Crivelli, C. K. Cesar, U. Gendotti, Canadian J. of Physics, 1 January 2011
\bibitem{Fee} M. S. Fee {\it et al.}, Phys. Rev. Lett. {\bf 70} (1993) 1397
\bibitem{Alpha} A. E. Charman {\it et al.} (ALPHA Collaboration), Nature Communications {\bf 4}, article 1785  (2013)
\bibitem{GBAR} G. Chardin et al., CERN-SPSC-2011-029 / SPSC-P-342 30/09/2011
\bibitem{Greaves} R. G. Greaves, M. D. Tinkle, and C. M. Surko,  Phys. Plasmas {\bf 1} (1994) 1439
\bibitem{Mariazzi} S. Mariazzi, P. Bettotti, R.S. Brusa, Phys. Rev. Lett. {\bf 104} (2010) 243401
\bibitem{Charlton} M. Charlton, Phys. Lett. {\bf A 143} (1990) 143
\bibitem{GemmaOki} G. Testera {\it et al.}, AIP Conf. Proc. {\bf 1037} (2008) 5
\bibitem{Merkt} E. Vliegen and F. Merkt, Phys. Rev. Lett {\bf 97} (2006) 033002
\bibitem{Oberthaler} M. K. Oberthaler {\it et al.}, Phys. Rev. {\bf A 54} (1996) 3165
\bibitem{OPERA} R. Acquafredda {\it et al.}, JINST {\bf 4} (2009) P04018.
 \bibitem{Aoki} S. Aoki {\it et al.}, Nucl. Instr. and Meth. in Phys. Res. {\bf A 488} (2002) 144
 \bibitem{Emul} C. Amsler {\it et al.}, J. of Instrumentation {\bf 8} (2013) P02015
\bibitem{Kimura} M. Kimura {\it et al.}, Nucl. Instr. and Meth.  in Phys. Res. {\bf A 711} (2013) 1
\bibitem{Lellis} G. de Lellis, A. Ereditato, K. Niwa, \emph{Nuclear Emulsions}, C. W. Fabjan and H. Schopper eds., Springer Materials, Landolt-B\"ornstein Database (http://www.springermaterials.com),
Springer-Verlag, Heidelberg, 2011 
\bibitem{Vienna} M. Kimura {\it et al.} (AEgIS Collaboration),  Nucl. Instr. and Meth. in Phys. Res {\bf A}, doi:10.1016/j.nima.2013.04.082
\bibitem{Nakamura} T. Nakamura {\it et al.}, Nucl. Instr. and Meth. in Phys. Res. {\bf A 556} (2006) 80.
\bibitem{Nakathesis} T. Nakamura, PhD Thesis, University of Nagoya (2005) 
\bibitem{CHIPSref}  P. V. Degtyarenko {\it et al.}, Eur. Phys. J. {\bf A 8} (2000) 217;\\
\hspace{1mm} P. V. Degtyarenko {\it et al.}, Eur. Phys. J. {\bf A 9} (2000) 411;\\
\hspace{1mm} P. V. Degtyarenko {\it et al.}, Eur. Phys. J. {\bf A 9} (2000) 421;\\
\hspace{1mm} M. Kossov, IEEE Trans. Nucl. Sci. {\bf 52} (2005) 2832
\bibitem{FTFPref} B. Andersson, G. Gustafson and B. Nilsson-Almqvist, Nucl. Phys. {\bf B 281} (1987) 289;\\
\hspace{1mm} B. Nilsson-Almquist and E. Stenlund, Comp. Phys. Comm. {\bf 43} (1987) 387
\bibitem{Galoyan} A. Galoyan and V. Uzhinsky, prep. arXiv:1208.3614 (2012)
\bibitem{Bendi} G. Bendiscioli and D. Kharzeev, Rev. Nuovo Cim. {\bf 17} (1994) 1
\bibitem{mini} AEgIS Collaboration, in preparation
\bibitem{cuda} NVIDIA homepage: http://www.nvidia.com/cuda
\bibitem{Smart} V. Smart {\it et al.}, Acta Phys. Polonica {\bf B 17} (1986) 42
\bibitem{EMgroup} CERN Emulsion Group, 5th Int. Conf. on Nuclear Photography, Geneva, Sept. 1964, pp.IV/1-4

\end{thebibliography}
\end{document}